\documentclass[%
 reprint,
superscriptaddress,
 amsmath,amssymb,
 aps
pra,
]{revtex4-2}

\usepackage{graphicx}
\usepackage{graphics}
\usepackage{dcolumn}
\usepackage{bm}
\usepackage{subcaption}
\usepackage{float}
\usepackage{xcolor}

\begin{document}

\preprint{APS/123-QED}

\title{Chaos in Optomechanical Systems coupled to a Non-Markovian environment}

\author{Pengju Chen}
\affiliation{%
Center for Quantum Science and Engineering\\
Department of Physics, Stevens Institute of Technology \\ Hoboken, New Jersey 07030, USA.
}%
\author{Nan Yang}
\affiliation{%
Center for Quantum Science and Engineering\\
Department of Physics, Stevens Institute of Technology \\ Hoboken, New Jersey 07030, USA.
}%
\author{Austen Couvertier}
\affiliation{%
Center for Quantum Science and Engineering\\
Department of Physics, Stevens Institute of Technology \\ Hoboken, New Jersey 07030, USA.
}%
\author{Quanzhen Ding}%
\affiliation{%
Center for Quantum Science and Engineering\\
Department of Physics, Stevens Institute of Technology \\ Hoboken, New Jersey 07030, USA.
}%
\author{Rupak Chatterjee}
\affiliation{%
Deptartment of Applied Physics, New York University Tandon School of Engineering \\ 
Brooklyn, New York 11201, USA.}%
\author{Ting Yu}\email{tyu1@stevens.edu}
\affiliation{%
 Center for Quantum Science and Engineering\\
 Department of Physics, Stevens Institute of Technology \\ Hoboken, New Jersey 07030, USA.
}%

\date{\today}

\begin{abstract}
We study the chaotic motion of a semi-classical optomechanical system coupled to a non-Markovian environment with a finite correlation time.
We show that the non-Markovian environment can significantly enhance chaos, by studying the emergence of chaos using Lyapunov exponent with the changing non-Markovian parameter.
It is observed that non-Markovian environment characterized by the Ornstein-Uhlenbeck type noise can modify the generation of chaos with different environmental memory times. 
As a comparison, the crossover properties from Markov to non-Markovian regimes are also discussed. Our findings indicate that the quantum memory effects on the onset of chaos may become a useful property to be investigated in quantum manipulations and control. \end{abstract}

\maketitle


\section{Introduction}


Optomechanics studies the interaction between light and mechanical systems \cite{aspelmeyer2014cavity,carmon2007chaotic}.
The nonlinear optical and mechanical interactions  have provided a new platform to realize quantum information processing \cite{yang2015noise,jiang2017chaos}, quantum sensing \cite{monifi2016optomechanically}, and to test the fundamental quantum physics such as decoherence \cite{larson2011photonic}, quantum entanglement \cite{wang2014nonlinear,yang2022macroscopic}, classical dynamical gauge
fields~\cite{walter2016classical}, and Parity-Time (PT) symmetry breaking \cite{lu2015p}, to name a few.  
In addition, many novel features of optomechanical systems involving chaotic properties have been investigated in various interesting settings~\cite{aspelmeyer2014cavity,carmon2007chaotic,sciamanna2016vibrations,navarro2017nonlinear,lee2009observation,sun2014chaotic,piazza2015self,zhang2020intermittent,yang2015noise,jiang2017chaos,monifi2016optomechanically,larson2011photonic,wang2014nonlinear,lu2015p,walter2016classical,wang2016transient},
one of the interesting physical problems in employing optomechanical systems is the chaotic motion in an  optomechanical system coupled to a dissipative environment \cite{bakemeier2015route}.
Such study has opened up new opportunities in studying the onset of chaos dynamics emerged from the non-linearity of the optomechanical systems in a semi-classical regime, and for a Markov environment, it has shown several interesting features on how the system parameters such as the detuning,  the optical and mechanical oscillator coupling and the external pumping power can affect the chaotic motion. 
This optomechanical platform has shed a new light into the study of chaotic motion induced by quantum systems which has attracted a wide spectrum of interests over the last decades \cite{nakamura1993new,heller1984bound,heller2018semiclassical,gutzwiller1990chaos,ullmo2008many,wright2010new,garcia2022out,novotny2023relative,roque2020nonlinear}. 

The sensitivity of chaotic motion to the parameters of the optomechanical system raises an interesting question: are there any other physical parameters that can play an important role in altering the regular and
chaotic motion of an open optomechanical system? One direct answer to this question is to go beyond the assumption of Markov environment. The purpose of this paper is to study the regular and chaotic motion of a nonlinear optomechanical systems coupled to a non-Markovian environment \cite{breuer2002theory,weiss2012quantum,de2017dynamics}. 
It is observed that a non-Markovian environment characterized by an Ornstein-Uhlenbeck type noise can modify chaotic motion with different environmental memory times. 
The advantages of choosing an Ornstein-Uhlenbeck type noise is that the environmental memory times may be conveniently characterized by a single parameter $\gamma$ that will provide a useful model to study the crossover properties from non-Markovian to Markov regimes ($ \gamma \rightarrow \infty$).  
Moreover,  our model has shown that the environmental memory time defined in this model may become a useful property to be investigated in quantum manipulations and control \cite{miki2023generating,liu2023phase}. More specifically, we will study the dynamics of the optomechanical system coupled to a non-Markovian environment and explore emergence of chaotic motion by using Lyapunov exponent with the varied environmental memory times \cite{strunz1999open,yu1999non,diosi1998non,strunz2004convolutionless,yu2004non,jing2010non,yang2012nonadiabatic,chen2014exact,xu2014perturbation}. As a result, our non-Markovian approach also provides a convenient pathway to the well-known Markov limit~\cite{dalibard1992wave,gisin1992quantum,carmichael1993open,wiseman1993interpretation,plenio1998quantum}.

The paper is organized as follows: in Section II, we consider the combined mechanical and optical modes coupled to two heat baths. 
By employing the approximate non-Markovian master equation, 
one can effectively simulate the dynamics of the optomechanical system. In Section III, we study the chaotic dynamics of the optomechanical system embedded in a non-Markovian environment. 
These results are used to illustrate how the environmental memory affects the onset of the chaos in various parameter settings. 
As a comparison, the crossover properties from Markovian to non-Markovian regimes are also discussed.
In Section IV,  we make some comments on several physical issues that are relevant to the discussions presented in this paper and further extension of the present study. Technical details are relegated to the appendix.

\begin{figure}
  \centering
  \includegraphics[width=6cm]{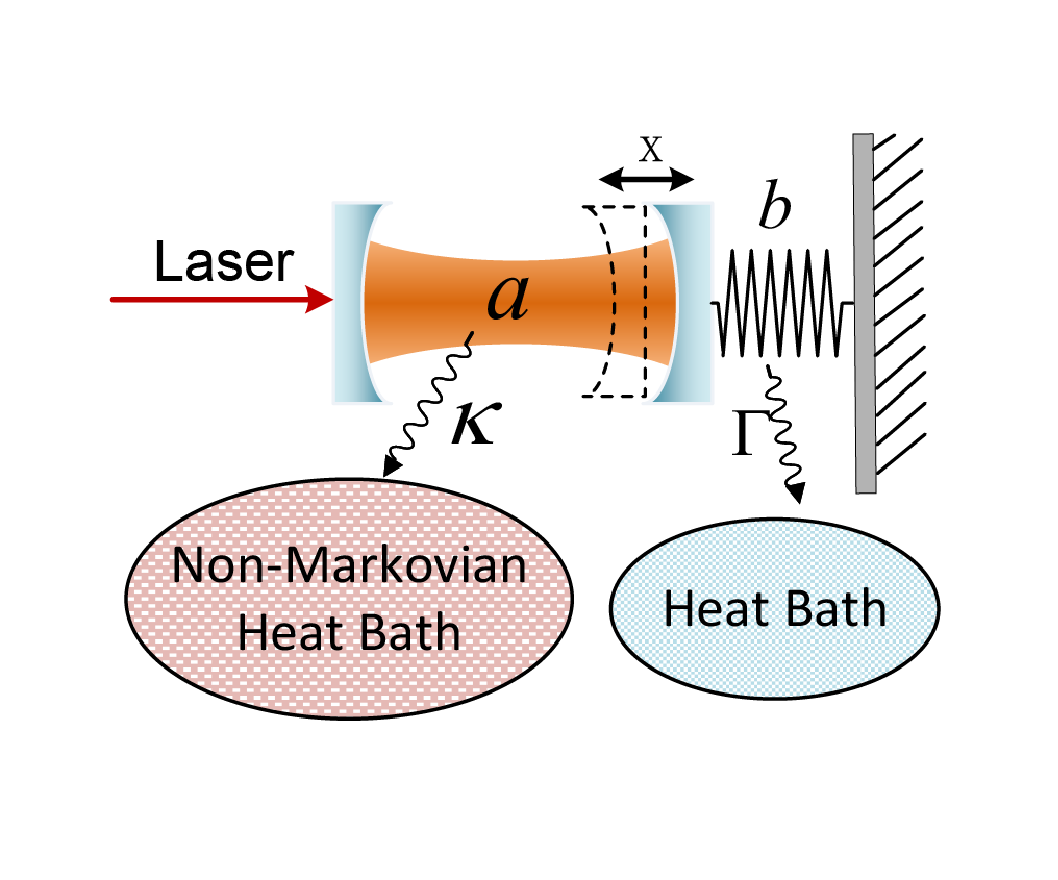}
  \caption{Schematic diagram of the optomechanical system, where the light field is considered to be in non-Markovian environment. The mechanical bath is set to be Markov to account for mechanical loss.}
  \label{fig1}
\end{figure}

\section{Optomechanical model coupled to a non-Markovian environment}
\label{model}
In this section, we consider an optomechanical system coupled to a non-Markovian environment.  We will use this model to investigate the interrelationship between the chaos and the environmental memory.
The optomechanical system consists of a mechanical mode couples with an optical mode, as exhibited in Figure~\ref{fig1}.

The optomechanical system under consideration is described by the Hamiltonian \cite{aspelmeyer2014cavity} (setting $\hbar=1$)
\begin{equation}\label{Hamiltonian}
H_{\rm s}
=\big[-\Delta_\omega+g_{0}(b+b^{\dagger})\big]a^{\dagger} a+\Omega b^{\dagger} b
+\alpha _{\textit{\texttt{L}}}(a^{\dagger}+a),
\end{equation}
where $a$ $(a^\dag)$ and $b $ $(b^\dag)$ are the annihilation (creation) operators of the optical mode and mechanical mode which satisfy the commutation relations $[a,a^\dagger]=1$ and $[b,b^\dagger]=1$, respectively.
Here $\Omega$ is the mechanical cantilever frequency, and $\Delta_\omega$ is the detuning parameter, which is
defined as $\Delta_\omega\equiv\omega_L-\omega_\mathrm{cav}$, 
where $\omega_L$ is the pumping laser frequency and $\omega_\mathrm{cav}$ is the resonant frequency of the cavity mode $a$.
$\alpha _L$ is the pumping laser amplitude and $g_{0}$ is the coupling constant between the two  mechanical and optical modes.

Since the optomechanical system is modeled as an open system, the cavity has radiative loss and cantilever has mechanical damping with respect to their local environments.
The two baths are assumed to be Bosonic, with the optical bath is set to be non-Markovian, which is the main focus of our study and may be adjusted experimentally \cite{liu2011experimental}.
While the mechanical bath is considered Markov to account for mechanical loss.
The damping parameters of the optical and mechanical baths are, respectively, $\kappa/\Omega=1$ and $\Gamma/\Omega=10^{-3}$.
Here the mutual correlation between two environments is omitted as the their interaction strength is much weaker than that between the systems and environments. 

Using the first-order, so-called post-Markov approximation of the non-Markovian quantum state diffusion (NMQSD) equation\cite{yu1999non}, the master equation takes the following form (details can be seen in Appendix \ref{deNMQSD})
\begin{widetext}
\begin{equation}\label{master}
\dot{\rho}
=
-i[H_s,\rho]+\Gamma D\big[b,\rho\big]
+\bigg\{f_{0}(t)[a\rho, a^{\dagger}]+i f_{1}(t)[a^{\dagger},[H_s,a]\rho]
+f_{2}(t)[a^{\dagger},[a^{\dagger},a]a\rho]+H.C. \bigg \},
\end{equation}
\end{widetext}
where $H.C.$ stands for Hermitian conjugate.
The so-called Lindblad term $D\big[b,\rho\big]=\big\{b\rho b^\dagger - \frac{1}{2}(b^\dagger b\rho+\rho b^\dagger b) \big\}$ represents the effect of the Markov mechanical bath.
And the non-Markovian environment effects represented by  the time-dependent coefficients  $f_0$, $f_1$, $f_2$  are given by
\begin{equation}
\begin{split}
&f_{0}(t)
=\int_{0}^{t}\alpha (t,s) \; ds,
\\
&f_{1}(t)
=\int_{0}^{t}\alpha (t,s) (t-s) \; ds,
\\
&f_{2}(t)
=\int_{0}^{t}\int_{0}^{s}\alpha (t,s) \alpha (s,u) (t-s) \; du \; ds.
\end{split}
\end{equation}
Where $\alpha(t,s)$ is the Ornstein-Uhlenbeck (O-U) type correlation function,
\begin{equation}\label{OU}
\alpha(t,s)=\frac{\kappa\gamma}{2}e^{-\gamma|t-s|},
\end{equation}
where $\tau_{\rm env}=1/\gamma$ is the environmental memory time, and $\kappa$ is the optical damping rate where $\kappa/\Omega=1$.

For convenience, we express the system parameters in units of $\Omega$ and introduce the dimensionless time parameter $\tau=\Omega t$.
Additionally, we introduce a dimensionless parameter \cite{marquardt2006dynamical,ludwig2008optomechanical,bakemeier2015route}
\begin{equation}
 P = \frac{8\alpha_L^2g_0^2}{\Omega^4}.
\end{equation}
The pumping parameter $P$ represents the strength of the laser input of the cavity.

We use the re-scaled mean values of the creation and annihilation operators $\alpha_1 = [\Omega/(2\alpha_L)]\langle a\rangle$, $\beta_1 = (g_0/\Omega)\langle b\rangle$ to represent photon and phonon modes.
In the bad-cavity limit, namely, the leakage rate $\kappa$  of the cavity is sufficiently high, such that $g_0/\kappa\ll 1$,   the semi-classical (SC) equations of motion are give by (the equations governing $\alpha_1^*$ and $\beta_1^*$ may be obtained easily),
\begin{equation}\label{first2}
\begin{split}
\frac{d\alpha_1}{d\tau} =& -i(1+f_1)\bigg\{(\beta_1+\beta_1^\ast)\alpha_1-\frac{\Delta_\omega}{\Omega}\alpha_1+\frac{1}{2}\bigg\}-\frac{f_0+f_2}{\Omega}\alpha_1,
\\
\frac{d\beta_1}{d\tau} =& -i\bigg\{\frac{P}{2}|\alpha_1|^2+\beta_1\bigg\}-\frac{\Gamma}{2\Omega}\beta_1,
\end{split}
\end{equation}
where we used the SC approximation $\langle (b^\dagger+b)a\rangle \approx \langle b^\dagger\rangle \langle a\rangle+\langle b\rangle\langle a\rangle$, which ignores all photon-phonon correlations.
The coefficients $f_0$, $f_1$ and $f_2$ are time-dependent and encode the information about the non-Markovian environment.

\section{Chaos in non-Markovian environments}
\label{chaosnm}

Effects of the  non-Markovian environment on the chaotic motion of the optomechanical system is the focus of this section. 
The optomechanical systems have proved their sensitivity to parameter changes of the input power $P$ and the detuning $\Delta_\omega$ in a Markov environment~\cite{bakemeier2015route}.  The new elements brought into the nonlinear equations of motion (\ref{first2}) are the time-dependent coefficients, one would expect that the chaotic motions should be modified by introducing more variable parameters.
The simulation methods used below allow us to explore how  the non-Markovian memory time influences the chaos generation in the optomechanical system. 

Our simulations are mainly based on observing the optomechanical system while changing the memory time of the optical bath represented by the parameter $\gamma$ (the inverse of memory time).
Furthermore, we vary the detuning $\Delta_\omega$ and the pumping strength $P$ to get a comprehensive picture of the chaos distribution of our system.
The initial states are set to be the vacuum states.
We use the maximal Lyapunov exponent ($\bm{LE}$) as the indicator of chaos,
which is calculated using the Wolf's method of phase reconstruction \cite{wolf1985determining} at the long term limit and excludes initial transient stage.
The positive $\bm{LE}$ implies the onset of chaos.

\subsection{Simulation results}
First, the phase space description provides a powerful tool to study chaos in Markov and non-Markovian environments. 
By taking the real and imaginary parts of the time series of $\alpha_1$ and $\beta_1$ from Eq.~(\ref{first2}), we can generate the four-dimensional phase diagrams for different memory times (the inverse of $\gamma$). 

\begin{figure*}
\centering
\begin{subfigure}{.3\textwidth}
  \centering
  \includegraphics[width=5cm]{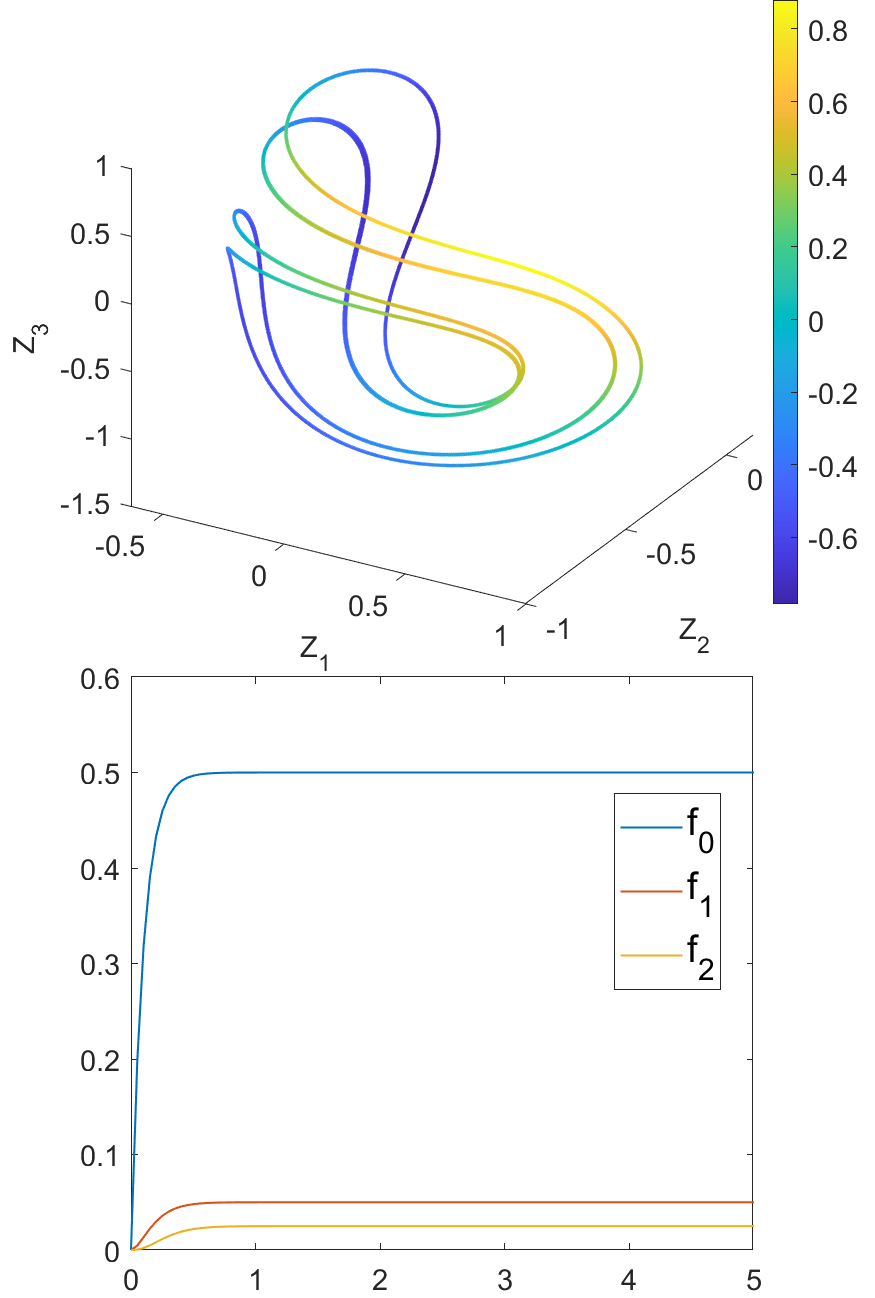}
  \caption{$\gamma = 10$, $\bm{LE}=0.0000196$.}
  \label{Phase_f_gamma_10}
\end{subfigure}%
\begin{subfigure}{.3\textwidth}
  \centering
  \includegraphics[width=5cm]{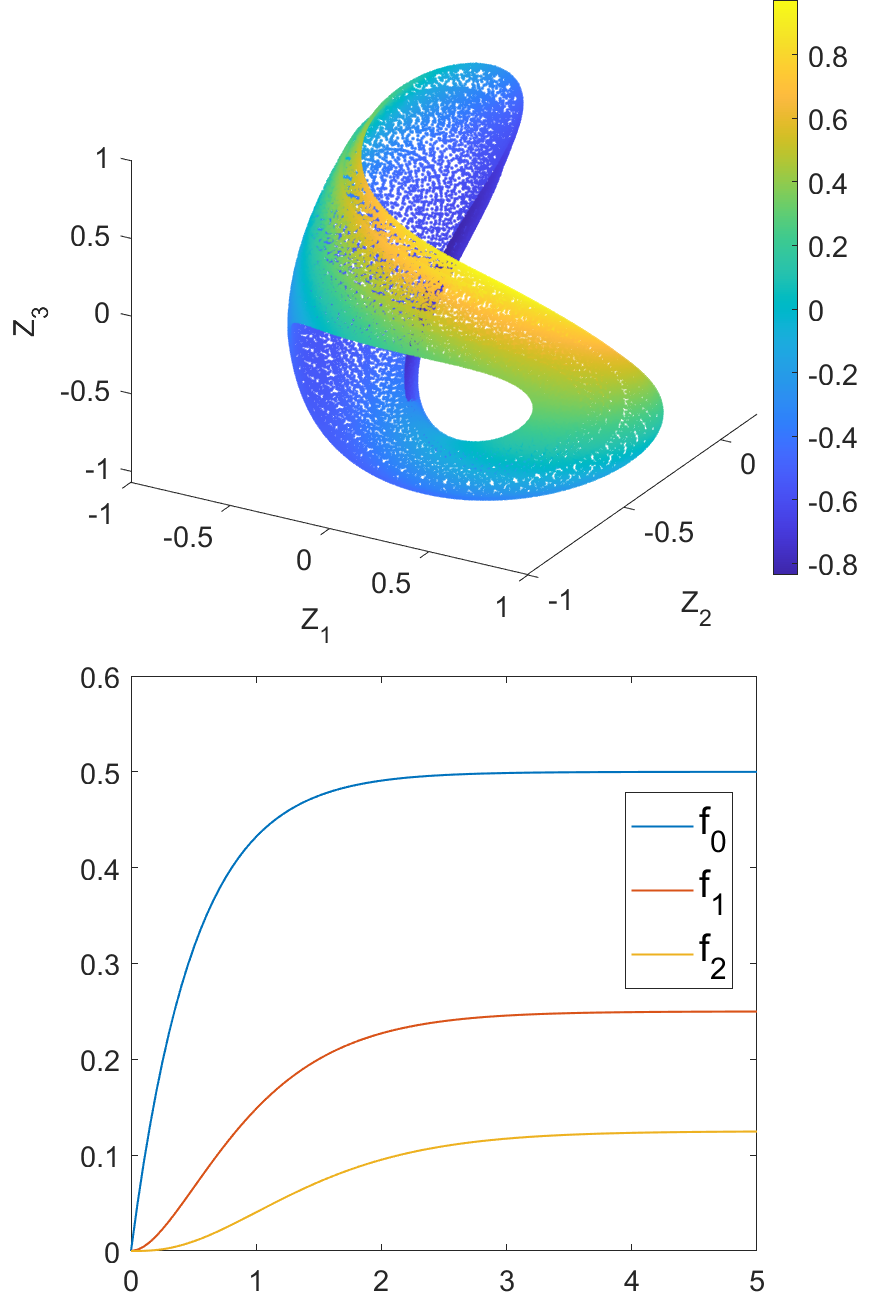}
  \caption{$\gamma = 2$, $\bm{LE}=0.1142$.}
  \label{Phase_f_gamma_2}
\end{subfigure}%
    \caption{The top 2 figures are phase diagrams with different memory times at $\Delta_\omega=-0.65$, $P=1.37$, such a point is chosen based on the bifurcation diagrams (Fig. \ref{BifurcationSC}). The coordinates $(Z_1,Z_2,Z_3,Z_4)$ are the real and imaginary parts of $\alpha_1$ and $\beta_1$. The 4th coordinate $Z_4$ is represented by scaled colours. As memory time increases ($\gamma$ decreases), the dynamics goes from regular to chaotic, indicated by both the phase diagrams and $\bm{LE}$. 
The bottom 2 figures show the evolution of $f_0(t)$, $f_1(t)$ and $f_2(t)$ with the integration time $\tau=5$. The decrease of $\gamma$ not only slows down the convergence of $f_i$ functions, it also changes the stable values of $f_1$ and $f_2$. }
    \label{SCPhase}
\end{figure*}

\begin{figure*}
\centering
\begin{subfigure}{.5\textwidth}
  \centering
  \includegraphics[width=8cm]{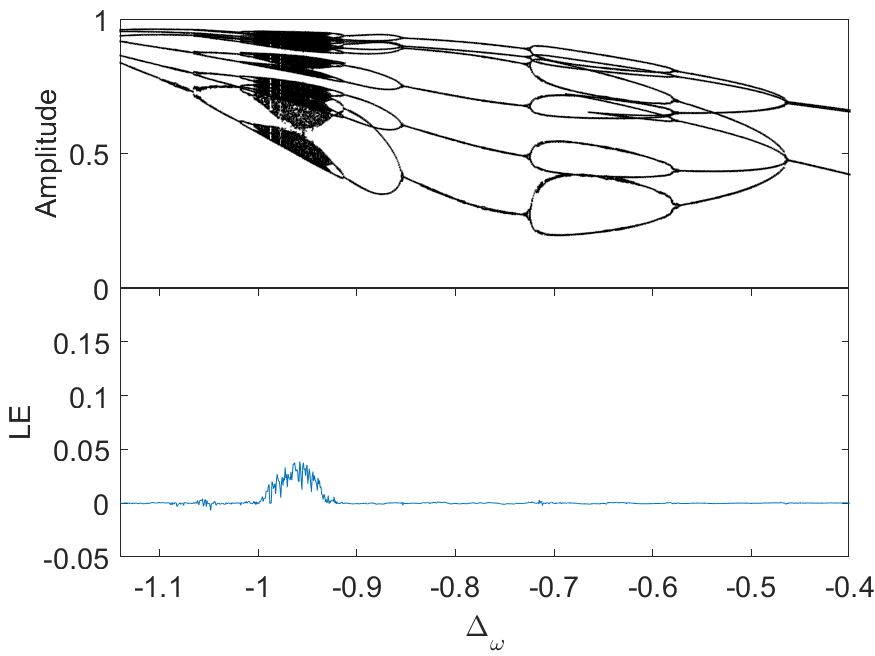}
  \caption{$\gamma = 10$}
  \label{Bifurcation_LE_gamma_10_aL_4.15}
\end{subfigure}%
\begin{subfigure}{.5\textwidth}
  \centering
  \includegraphics[width=8cm]{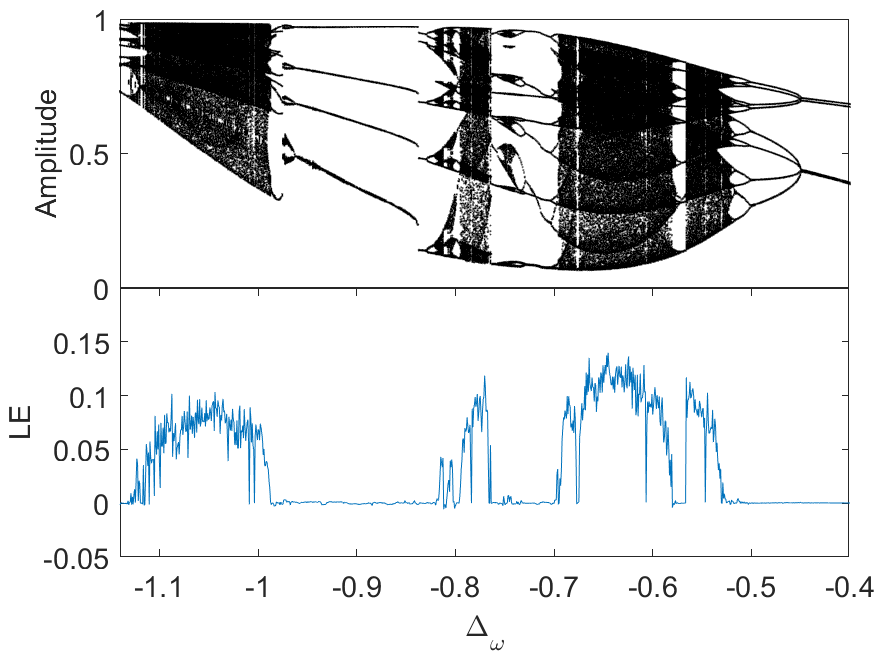}
  \caption{$\gamma = 2$}
  \label{Bifurcation_LE_gamma_2_aL_4.15}
\end{subfigure}
\caption{Bifurcation diagrams and the corresponding maximum Lyapunov exponents ($\bm{LE}$) with different memory times at $P = 1.37$. In (a), the chaotic region is bounded within a small segment $\Delta_\omega \in [-1.03, -0.92]$. In (b), one could see more bifurcation and the chaotic region expands to $\Delta_\omega \in [-1.13, -0.99]$ and a new chaotic region $\Delta_\omega \in [-0.83, -0.52]$ emerges, with some inter-adjacent regular regions inside.}
\label{BifurcationSC}
\end{figure*}

Figure~(\ref{SCPhase}) illustrates the system dynamics under different memory times for a chosen point ($P=1.37$ and $\Delta_\omega=-0.65$), such a point is chosen based on the bifurcation diagrams (Fig. \ref{BifurcationSC}), where a non-chaotic point when $\gamma=10$ ($\bm{LE}=0.0000196$) becomes chaotic when $\gamma=2$ ($\bm{LE}=0.1142$), as indicated by both phase diagrams and $\bm{LE}$. 
It is observed that the increase of memory time (decrease of $\gamma$) results in the system changing from regular to chaotic, such a transition is caused by the change of evolutions of $f_0(t)$, $f_1(t)$ and $f_2(t)$ (simply denoted as $f_i$), as is shown in Fig.~(\ref{SCPhase}). 
Note that we take the integration time of $f_i$ for just $\tau=5$, since they converge very fast, and the integration time for the phase diagrams is $\tau=10000$ to accommodate for chaos.
The increase of memory time slows down the convergence of $f_i$ functions while changing the stable values of $f_1$ and $f_2$.

\begin{figure*}
\centering
\begin{subfigure}{.4\textwidth}
  \centering
  \includegraphics[width=8cm]{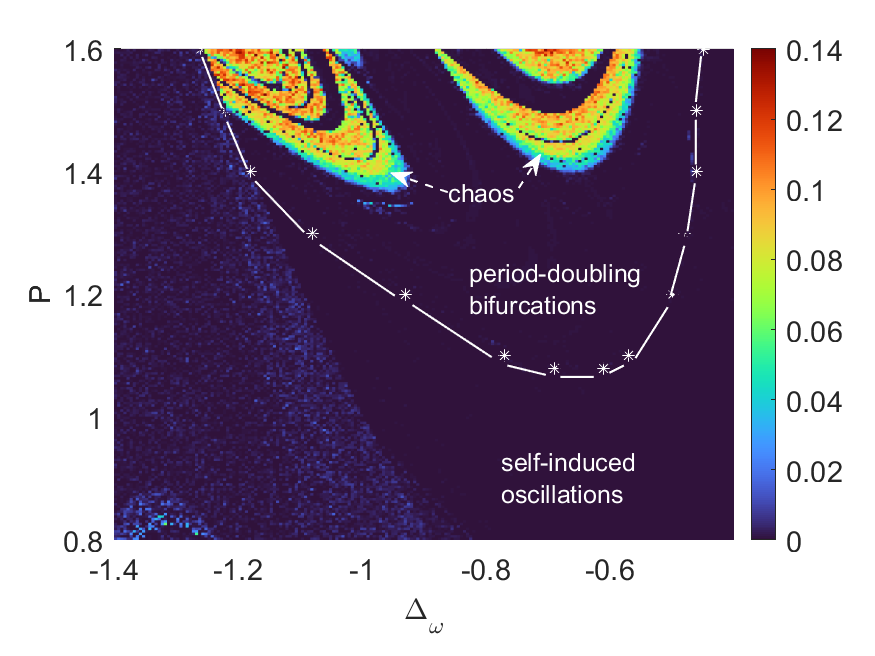}
  \caption{$\gamma=10$}
  \label{LE_parfor_10}
\end{subfigure}
\hspace{1cm}
\begin{subfigure}{.4\textwidth}
  \centering
  \includegraphics[width=8cm]{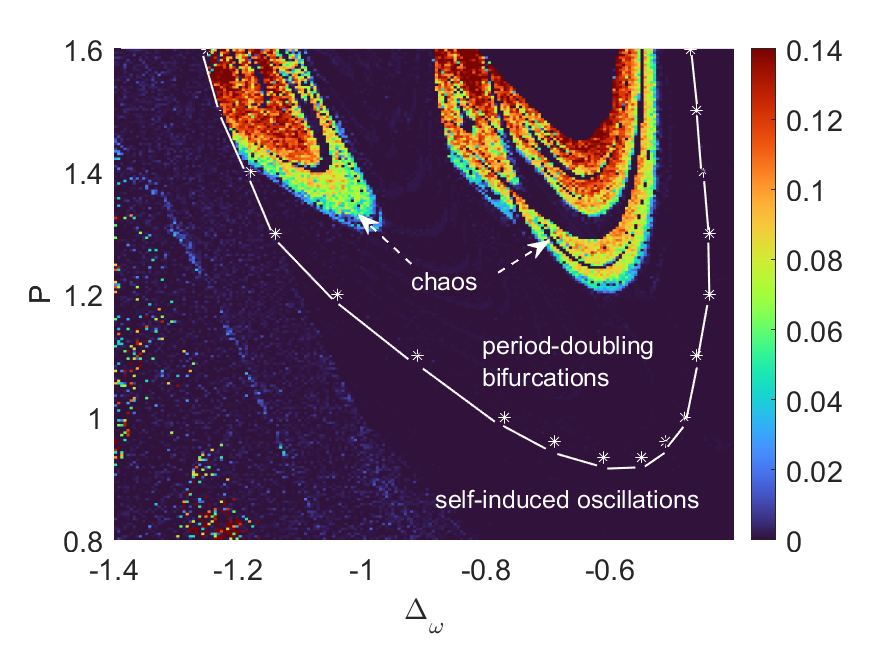}
  \caption{$\gamma=2$}
  \label{LE_parfor_1}
\end{subfigure}
  \caption{Pictures of chaotic regions of the optomechanical systems with different memory times plotted in the $P$-$\Delta_\omega$ plane. The colour scale shows the value of maximal Lyapunov exponent ($\bm{LE}$) of every data point. White stars represent data points extracted from bifurcation diagrams, while dashed lines interpolate between the data. Comparing \ref{LE_parfor_10} and \ref{LE_parfor_1}, the chaotic area expands as the memory time gets longer ($\gamma$ decreases).}
  \label{LE_parfor}
\end{figure*}
\begin{figure*}
\centering
  \includegraphics[width=8cm]{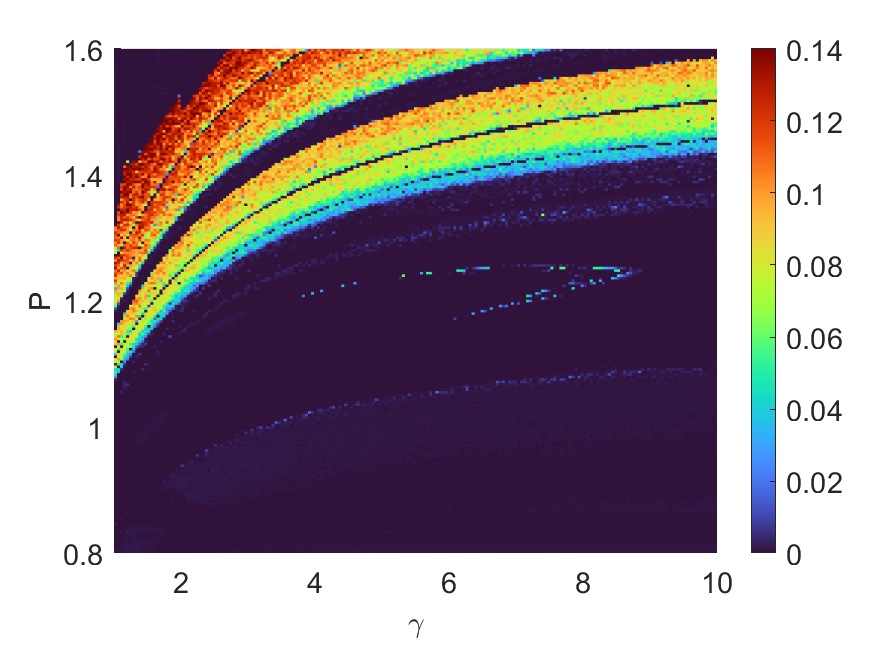}
  \caption{Fixing $\Delta_\omega = -0.6$, the plot shows the chaos landscape of $\gamma$ vs $P$. The pumping bar of chaos generation decreases as memory time increases ($\gamma$ decreases). Also, each $\gamma$ corresponds to a specific vertical line of chaos distribution, which could be exploited as a tool of measuring an unknown material's memory time.}
  \label{LE_gamma_D_-0.6}
\end{figure*}

Next, we plot the bifurcation diagrams of the cantilever oscillation.
Fixing $P = 1.37$,
at $\gamma=10$, the period-doubling bifurcation takes place, as is exhibited in Fig.~(\ref{Bifurcation_LE_gamma_10_aL_4.15}), where
the chaotic region is bounded within a small segment $\Delta_\omega \in [-1.03, -0.92]$.
With the same $P$, we change $\gamma$ to 2.
Fig.~(\ref{Bifurcation_LE_gamma_2_aL_4.15}) shows more bifurcations, and the chaotic region expanding to $\Delta_\omega \in [-1.13, -0.99]$ while a new chaotic region at $\Delta_\omega \in [-0.83, -0.52]$ emerges, with some inter-adjacent regular regions inside.
The comparison between Fig.~(\ref{Bifurcation_LE_gamma_10_aL_4.15}) and (\ref{Bifurcation_LE_gamma_2_aL_4.15}) indicates that the longer memory time expand chaotic regions. Physically, this is a very interesting 
observation. It indicates that an engineered environment with finite correlation times may be used to control the onset of chaos and system dynamics. Such an idea may be of interest in quantum measurement and
control.
Note that $\gamma=2$ is not far away from the Markov limit, but it still causes significant change to chaotic dynamics, further indicating that the chaos is very sensitive to parameter changes, including the memory time.

To summarize the appearances of chaos, the $\bm{LE}$ of every data point is plotted to form
global chaos landscapes, as is shown in Fig.~(\ref{LE_parfor_10}) for $\gamma=10$ and (\ref{LE_parfor_1}) for $\gamma=2$.
The parameter ranges are set to be $P\in [0.8, 1.6]$ and $\Delta_\omega\in [-1.4, -0.4]$.
As the pumping strength increases, we see more chaotic motion \cite{bakemeier2015route,ludwig2008optomechanical,marquardt2006dynamical}.
In both cases, the chaotic regions have shown some fine structures of inter-adjacent regular regions.
For our model, we have shown that the non-Markovian environment may expand the chaotic regions
while lowering the pumping energy needed for chaos generation.

The chaos landscape based on $\gamma$ vs $P$ can be seen in Fig.~(\ref{LE_gamma_D_-0.6}), with $\Delta_\omega = -0.6$.
The pumping bar of chaos generation decreases as memory time increases ($\gamma$ decreases).
Each $\gamma$ corresponds to a specific vertical line of chaos distribution, due to the chaos' sensitivity to parameters.
The uniqueness of the chaos distribution regarding each specific $\gamma$ could be the tool of ultra-sensitive measurement of memory time of an unknown material.
           
We have seen that the increase of memory time can expand chaotic areas in the parameter plane and decrease the pumping threshold 
for chaos generation.The non-Markovian environment has modified the dissipative rate (time-dependent), as such it has modified the dynamics
of chaotic motion.
Then non-Markovian properties of the environment has brought about several new features such as lowering the pumping threshold while expanding chaotic regions across the map.
It is noted that a similar phenomenon was also observed that the environment memory may enhance entanglement generation in non-Markovian regime \cite{mu2016memory}. 

\subsection{The comparison between Markov and non-Markovian regimes}
How is the chaotic motion in the optomechanical system coupled to a non-Markovian environment?  There is no single prescription for the chaos behaviors because the onset of chaos is sensitive to
the different physical parameters such as the detuning, pumping power and the environmental memory time. We now turn to the comparison of the Markov and non-Markovian regimes.
We find the post-Markov approximation particularly convenient for this purpose as the post-Markov regime provides the first-order correction to the well-known Markov regime.

\begin{figure*}
\centering
\begin{subfigure}{.5\textwidth}
  \centering
  \includegraphics[width=8cm]{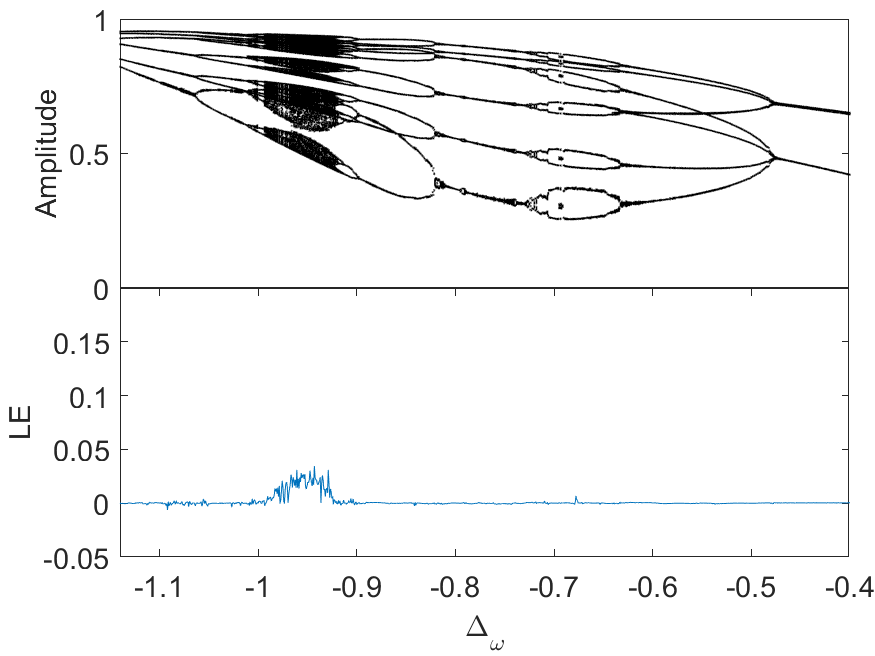}
  \caption{Markov}
  \label{Bifurcation_LE_Markov_P_1.4}
\label{Markov1}
\end{subfigure}%
\begin{subfigure}{.5\textwidth}
  \centering
  \includegraphics[width=8cm]{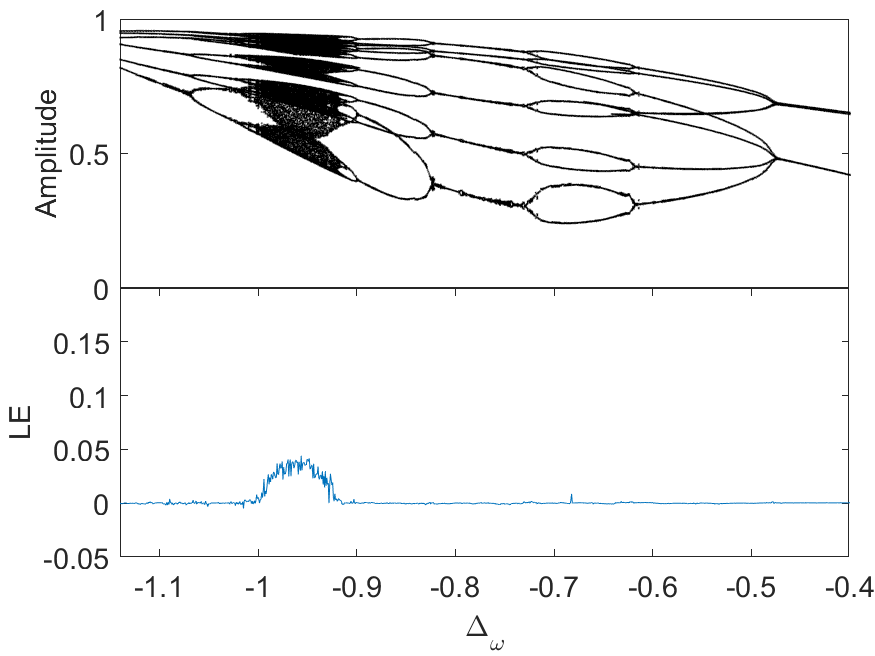}
  \caption{$\gamma = 100$}
  \label{Bifurcation_LE_gamma_100_P_1.4}
\label{NMarkov1}
\end{subfigure}
\caption{Bifurcation diagrams at $P = 1.4$. They show the transition of environment from Markovian regime to non-Markovian.}
\label{MarkovBifurCompare}
\end{figure*}

We begin by choosing the Markov limit (i.e., $\gamma \rightarrow \infty$),  the correction function is described by a delta function: $\alpha(t-s)= \kappa \delta(t-s)$, then $f_0=\kappa/2$,  $f_1=0$ and $f_2=0$,  substitution of these constant coefficients into Eq.~(\ref{first2})  and setting  $P=1.4$, we can numerically generate the bifurcation diagram and the corresponding $\bm{LE}$ of the Markov case (Fig.~ \ref{Markov1}).  
As comparison, we also plotted the same bifurcation diagram by using the non-Makovian equations with the parameter $\gamma=100$ (Fig.~6b). We note, however,  that, in the long-time limit ($t \rightarrow \infty$), 
the coefficients with $\gamma=100$ are $f_0=0.5$, $f_1=0.005$ and $f_2=0.0025$, though $f_1, f_2$ are small,  they are not zeros (as in the case of Markov approximation).  Our numerical simulations show
the sensitivity of chaotic motion on the presence of non-zero coefficients $f_1, f_2$. 
Clearly, as $\gamma$ becomes larger and larger, the distinction in the bifurcation diagrams will be smaller and smaller, approaching to the  Markov limit (Fig.~6a). 
We have numerically demonstrated that, as the environmental memory time $\tau_{\rm env}=1/\gamma$ decreases, the chaotic regions will shrink and converge to the Markov case.

\section{Conclusion}
The chaotic motion emerged from open quantum systems is important for a better understanding quantum open system dynamics, and it is also of great interest for many physical applications such as 
quantum measurement and quantum sensing. Based on the optomechanical system coupled to a non-Markovian environment described the finite correlation time $\tau_{\rm env}=1/\gamma$,  we have shown that the finite environmental memory times can affect 
chaos generation in various interesting parameter ranges.  It is seen that the environmental memory effect effectively enlarge the chaotic region while lowering the necessary pumping for chaos generation. 
Such a phenomenon provides an additional parameter to control chaotic motion in optomechanical systems \cite{harayama2011fast,shen2023harnessing}. 
There are many interesting questions remaining, such as the relationship between optomechanical entanglement generation and the chaos, the high-order non-Markovian effects on the chaos behaviors,
and given that non-Markovian process is time dependent, its interaction with the dynamics of out-of-time-ordered correlators (OTOCs)~\cite{sun2020out,chavez2019quantum,buijsman2017nonergodicity} as signature of quantum chaos could be extremely relevant in the system presented, these will be left for future publications.

\section*{Acknowledgement}
This work is partially supported by the ART020-Quantum Technologies Project. 
We thank Dr. Mengdi Sun, Kenneth Mui, and Yifan Shi for helpful discussions on optomechanics.
\section*{Disclosures} The authors declare no conflicts of interest.
\section*{Data availability} 
Data underlying the results presented in this paper is public and could be found in this github repository: https://github.com/DOGECHANN/Chaos-in-Non-Markovian-Optomechanical-Systems

\appendix
\section{The non-Markovian quantum state diffusion (NMQSD) and master equation}\label{deNMQSD}
For convenience of the general readership,
the following section provides details about the master equation used in this paper.

Both environments are set to be at zero temperature, and separate from each other.
Since the Markov environment is well-understood and its formation can be added to the master equation easily,
we can focus on the optical bath and use the non-Markovian quantum state diffusion (QSD) equation to derive the master equation.

The Bosonic bath for the optical mode can be described by
a set of harmonic oscillators using creation/annihilation operators $(c_j^\dagger , c_i)$ satisfying the commutation relation $[c_i, c_j^\dagger]=\delta_{ij}$, with the Hamiltonian $H_{B}=\sum_j \omega_j c^\dagger_j c_j$ \cite{breuer2002theory,weiss2012quantum,mu2016memory}.
The interaction Hamiltonian between the system and the optical bath
is described by
\begin{equation}
H_{I}
=\sum_j g_j (L c^\dagger_j +L^\dagger c_j),
\end{equation}
where
$L = a$ is the Lindblad operator representing the optical damping and
$g_j$ are the system-bath coupling strength \cite{mu2016memory}.
We have taken the rotating-wave approximation (RWA) for our formalism as we have only considered weak-coupling strengths, high cavity leakage rate and weakly non-Markovian regimes (for a systematic analysis of the RWA, see, ~\cite{scala2021beyond,burgarth2023taming}).


Making use of the non-Markovian quantum state diffusion (NMQSD) \cite{diosi1998non,strunz1999open,mu2016memory}, while assuming that the system is initially uncorrelated with the environment
\begin{equation}\label{NMQSD1}
\partial_t |\psi_t(z^\ast)\rangle
=\bigg[-iH_s+az^\ast_t-a^\dagger \overline{O}(t,z^\ast)  \bigg]|\psi_t(z^\ast)\rangle,
\end{equation}
where $O(t,s,z^\ast)\psi_t\equiv\frac{\delta\psi_t}{\delta z_s}$ and $\overline{O}(t,z^\ast)\equiv\int^t_0ds\alpha(t,s)O(t,s,z^\ast)$ with the initial condition $O(t, s=t, z^\ast) = a$. 
$\alpha(t,s)=\frac{\kappa\gamma}{2}e^{-\gamma|t-s|}$ is the Ornstein-Uhlenbeck (O-U) correlation function.
$z_t^\ast = -i\sum_jg_jz^\ast_je^{i\omega_jt}$ is a colored complex Gaussian process satisfying
\begin{equation}
M[z_t^\ast z_s]=\alpha(t,s).
\
M[z_t z_s]=0,
\end{equation}
where $M[\cdot]\equiv\int\frac{dz^2}{\pi}e^{-|z|^2}[\cdot]$ denotes the ensemble average over the noise $z_t$.

To solve (\ref{NMQSD1}), one needs to find the operator $O(t,s,z^\ast)$.
Under the condition that the memory time is not too long,
we expand $O(t,s,z^\ast)$ in powers of $(t-s)$  \cite{yu1999non}
\begin{equation}\label{Oex}
O(t,s,z^\ast)=O(s,s,z^\ast)+\frac{\partial O(t,s,z^\ast)}{\partial t}\bigg|_{t=s}(t-s)+\cdots,
\end{equation}
which is a systematic expansion of the non-Markovian QSD.
The zeroth-order term corresponds to the standard Markov QSD when $1/\gamma\to 0$.
We choose the first-order approximation of the operator $O(t,s,z^\ast)$ since the memory time is assumed to be short.

At the time point $t=s$, the expression of the operator $O(t,s,z^\ast)$ is given by \cite{yu1999non}
\begin{align}
O(s,s,z^\ast) &= a,
\\
\frac{\partial O(t,s,z^\ast)}{\partial t}\bigg|_{t=s}&=-i[H_s,a]-\int^s_0\alpha(s,u)du[a^\dagger,a]a,
\end{align}
where $H_s$ is the system Hamiltonian and $a$ is the Lindblad operator.
At this time point, the operator $\overline{O}$ becomes
\begin{equation}\label{Obar}
\overline{O} = f_0(t)a-f_1(t)i[H_s,a]-f_2(t)[a^\dagger,a]a,
\end{equation}
where
\begin{equation}
\begin{split}
&f_{0}(t)
=\int_{0}^{t}\alpha (t,s) \; ds,
\\
&f_{1}(t)
=\int_{0}^{t}\alpha (t,s) (t-s) \; ds,
\\
&f_{2}(t)
=\int_{0}^{t}\int_{0}^{s}\alpha (t,s) \alpha (s,u) (t-s) \; du \; ds.
\end{split}
\end{equation}
The explicit form of $f_i$ is given by
\begin{align}
f_{0}(t)
&=\frac{\kappa}{2}(1-e^{-\gamma t}),
\\
f_{1}(t)
&=\frac{\kappa}{2\gamma}(1-e^{-\gamma t}-\gamma t e^{-\gamma t}),
\\
f_{2}(t)
&=\frac{\kappa^2}{4\gamma}\bigg(1-e^{-\gamma t}-\gamma t e^{-\gamma t}-\frac{1}{2}\gamma^2 t^2 e^{-\gamma t}\bigg).
\end{align}
One can turn (\ref{NMQSD1}) into a master equation by taking the ensemble mean over the noise $z_t$ and introducing the reduced density matrix
$\rho_t = M[|\psi_t(z^\ast)\rangle\langle\psi_t(z^\ast)|]$.
When the environment is not far away from Markov,
the dependence of the operator $\overline{O}(t,z^\ast)$ on the noise $z_t$ is negligible.
Under this approximation, the master equation \cite{yu1999non} takes the form
\begin{equation}
\frac{d}{dt}\rho_t = -i[H_s,\rho_t]+[a,\rho_t\overline{O}(t)^\dagger]-[a^\dagger,\overline{O}(t)\rho_t].
\end{equation}
For further details about the derivation of the master equation, see \cite{strunz2004convolutionless,chen2014exact}.

Since the mechanical bath is separate and Markov, it can be added to the master equation by simply including the appropriate Lindblad term which is given by
\begin{equation}
\Gamma D\big[b,\rho\big]=\Gamma\big\{b\rho b^\dagger - \frac{1}{2}(b^\dagger b\rho+\rho b^\dagger b) \big\},
\end{equation}
where $\Gamma/\Omega=10^{-3}$ is the mechanical damping.

Using the first order approximation of $\overline{O}(t)$ (\ref{Obar}), the master equation takes the final form
\begin{widetext}
\begin{equation}
\dot{\rho}
=
-i[H_s,\rho]+\Gamma D\big[b,\rho\big]
+\bigg\{f_{0}(t)[a\rho, a^{\dagger}]+i f_{1}(t)[a^{\dagger},[H_s,a]\rho]
+f_{2}(t)[a^{\dagger},[a^{\dagger},a]a\rho]+H.C. \bigg \}
,
\end{equation}
\end{widetext}
where $H.C.$ stands for Hermitian conjugate.

\bibliographystyle{apsrev4-2}
\bibliography{CNMO}

\providecommand{\noopsort}[1]{}\providecommand{\singleletter}[1]{#1}%
\begin{thebibliography}{58}%
\makeatletter
\providecommand \@ifxundefined [1]{%
 \@ifx{#1\undefined}
}%
\providecommand \@ifnum [1]{%
 \ifnum #1\expandafter \@firstoftwo
 \else \expandafter \@secondoftwo
 \fi
}%
\providecommand \@ifx [1]{%
 \ifx #1\expandafter \@firstoftwo
 \else \expandafter \@secondoftwo
 \fi
}%
\providecommand \natexlab [1]{#1}%
\providecommand \enquote  [1]{``#1''}%
\providecommand \bibnamefont  [1]{#1}%
\providecommand \bibfnamefont [1]{#1}%
\providecommand \citenamefont [1]{#1}%
\providecommand \href@noop [0]{\@secondoftwo}%
\providecommand \href [0]{\begingroup \@sanitize@url \@href}%
\providecommand \@href[1]{\@@startlink{#1}\@@href}%
\providecommand \@@href[1]{\endgroup#1\@@endlink}%
\providecommand \@sanitize@url [0]{\catcode `\\12\catcode `\$12\catcode `\&12\catcode `\#12\catcode `\^12\catcode `\_12\catcode `\%12\relax}%
\providecommand \@@startlink[1]{}%
\providecommand \@@endlink[0]{}%
\providecommand \url  [0]{\begingroup\@sanitize@url \@url }%
\providecommand \@url [1]{\endgroup\@href {#1}{\urlprefix }}%
\providecommand \urlprefix  [0]{URL }%
\providecommand \Eprint [0]{\href }%
\providecommand \doibase [0]{https://doi.org/}%
\providecommand \selectlanguage [0]{\@gobble}%
\providecommand \bibinfo  [0]{\@secondoftwo}%
\providecommand \bibfield  [0]{\@secondoftwo}%
\providecommand \translation [1]{[#1]}%
\providecommand \BibitemOpen [0]{}%
\providecommand \bibitemStop [0]{}%
\providecommand \bibitemNoStop [0]{.\EOS\space}%
\providecommand \EOS [0]{\spacefactor3000\relax}%
\providecommand \BibitemShut  [1]{\csname bibitem#1\endcsname}%
\let\auto@bib@innerbib\@empty
\bibitem [{\citenamefont {Aspelmeyer}\ \emph {et~al.}(2014)\citenamefont {Aspelmeyer}, \citenamefont {Kippenberg},\ and\ \citenamefont {Marquardt}}]{aspelmeyer2014cavity}%
  \BibitemOpen
  \bibfield  {author} {\bibinfo {author} {\bibfnamefont {M.}~\bibnamefont {Aspelmeyer}}, \bibinfo {author} {\bibfnamefont {T.~J.}\ \bibnamefont {Kippenberg}},\ and\ \bibinfo {author} {\bibfnamefont {F.}~\bibnamefont {Marquardt}},\ }\href@noop {} {\bibfield  {journal} {\bibinfo  {journal} {Rev. Mod. Phys.}\ }\textbf {\bibinfo {volume} {86}},\ \bibinfo {pages} {1391} (\bibinfo {year} {2014})}\BibitemShut {NoStop}%
\bibitem [{\citenamefont {Carmon}\ \emph {et~al.}(2007)\citenamefont {Carmon}, \citenamefont {Cross},\ and\ \citenamefont {Vahala}}]{carmon2007chaotic}%
  \BibitemOpen
  \bibfield  {author} {\bibinfo {author} {\bibfnamefont {T.}~\bibnamefont {Carmon}}, \bibinfo {author} {\bibfnamefont {M.~C.}\ \bibnamefont {Cross}},\ and\ \bibinfo {author} {\bibfnamefont {K.~J.}\ \bibnamefont {Vahala}},\ }\href@noop {} {\bibfield  {journal} {\bibinfo  {journal} {Phys. Rev. Lett.}\ }\textbf {\bibinfo {volume} {98}},\ \bibinfo {pages} {167203} (\bibinfo {year} {2007})}\BibitemShut {NoStop}%
\bibitem [{\citenamefont {Yang}\ \emph {et~al.}(2015)\citenamefont {Yang}, \citenamefont {Zhang}, \citenamefont {Wang}, \citenamefont {Liu}, \citenamefont {Wu}, \citenamefont {Liu}, \citenamefont {Li},\ and\ \citenamefont {Nori}}]{yang2015noise}%
  \BibitemOpen
  \bibfield  {author} {\bibinfo {author} {\bibfnamefont {N.}~\bibnamefont {Yang}}, \bibinfo {author} {\bibfnamefont {J.}~\bibnamefont {Zhang}}, \bibinfo {author} {\bibfnamefont {H.}~\bibnamefont {Wang}}, \bibinfo {author} {\bibfnamefont {Y.~X.}\ \bibnamefont {Liu}}, \bibinfo {author} {\bibfnamefont {R.~B.}\ \bibnamefont {Wu}}, \bibinfo {author} {\bibfnamefont {L.~Q.}\ \bibnamefont {Liu}}, \bibinfo {author} {\bibfnamefont {C.~W.}\ \bibnamefont {Li}},\ and\ \bibinfo {author} {\bibfnamefont {F.}~\bibnamefont {Nori}},\ }\href {https://doi.org/10.1103/PhysRevA.92.033812} {\bibfield  {journal} {\bibinfo  {journal} {Phys. Rev. A}\ }\textbf {\bibinfo {volume} {92}},\ \bibinfo {pages} {033812} (\bibinfo {year} {2015})}\BibitemShut {NoStop}%
\bibitem [{\citenamefont {Jiang}\ \emph {et~al.}(2017)\citenamefont {Jiang}, \citenamefont {Shao}, \citenamefont {Zhang}, \citenamefont {Yi}, \citenamefont {Wiersig}, \citenamefont {Wang}, \citenamefont {Gong}, \citenamefont {Loncar}, \citenamefont {Yang},\ and\ \citenamefont {Xiao}}]{jiang2017chaos}%
  \BibitemOpen
  \bibfield  {author} {\bibinfo {author} {\bibfnamefont {X.~F.}\ \bibnamefont {Jiang}}, \bibinfo {author} {\bibfnamefont {L.~B.}\ \bibnamefont {Shao}}, \bibinfo {author} {\bibfnamefont {S.~X.}\ \bibnamefont {Zhang}}, \bibinfo {author} {\bibfnamefont {X.}~\bibnamefont {Yi}}, \bibinfo {author} {\bibfnamefont {J.}~\bibnamefont {Wiersig}}, \bibinfo {author} {\bibfnamefont {L.}~\bibnamefont {Wang}}, \bibinfo {author} {\bibfnamefont {Q.~H.}\ \bibnamefont {Gong}}, \bibinfo {author} {\bibfnamefont {M.}~\bibnamefont {Loncar}}, \bibinfo {author} {\bibfnamefont {L.}~\bibnamefont {Yang}},\ and\ \bibinfo {author} {\bibfnamefont {Y.~F.}\ \bibnamefont {Xiao}},\ }\href@noop {} {\bibfield  {journal} {\bibinfo  {journal} {Science}\ }\textbf {\bibinfo {volume} {358}},\ \bibinfo {pages} {344} (\bibinfo {year} {2017})}\BibitemShut {NoStop}%
\bibitem [{\citenamefont {Monifi}\ \emph {et~al.}(2016)\citenamefont {Monifi}, \citenamefont {Zhang}, \citenamefont {{\"O}zdemir}, \citenamefont {Peng}, \citenamefont {Liu}, \citenamefont {Bo}, \citenamefont {Nori},\ and\ \citenamefont {Yang}}]{monifi2016optomechanically}%
  \BibitemOpen
  \bibfield  {author} {\bibinfo {author} {\bibfnamefont {F.}~\bibnamefont {Monifi}}, \bibinfo {author} {\bibfnamefont {J.}~\bibnamefont {Zhang}}, \bibinfo {author} {\bibfnamefont {S.~K.}\ \bibnamefont {{\"O}zdemir}}, \bibinfo {author} {\bibfnamefont {B.}~\bibnamefont {Peng}}, \bibinfo {author} {\bibfnamefont {Y.~X.}\ \bibnamefont {Liu}}, \bibinfo {author} {\bibfnamefont {F.}~\bibnamefont {Bo}}, \bibinfo {author} {\bibfnamefont {F.}~\bibnamefont {Nori}},\ and\ \bibinfo {author} {\bibfnamefont {L.}~\bibnamefont {Yang}},\ }\href@noop {} {\bibfield  {journal} {\bibinfo  {journal} {Nat. Photon.}\ }\textbf {\bibinfo {volume} {10}},\ \bibinfo {pages} {399} (\bibinfo {year} {2016})}\BibitemShut {NoStop}%
\bibitem [{\citenamefont {Larson}\ and\ \citenamefont {Horsdal}(2011)}]{larson2011photonic}%
  \BibitemOpen
  \bibfield  {author} {\bibinfo {author} {\bibfnamefont {J.}~\bibnamefont {Larson}}\ and\ \bibinfo {author} {\bibfnamefont {M.}~\bibnamefont {Horsdal}},\ }\href@noop {} {\bibfield  {journal} {\bibinfo  {journal} {Phys. Rev. A}\ }\textbf {\bibinfo {volume} {84}},\ \bibinfo {pages} {021804(R)} (\bibinfo {year} {2011})}\BibitemShut {NoStop}%
\bibitem [{\citenamefont {Wang}\ \emph {et~al.}(2014)\citenamefont {Wang}, \citenamefont {Huang}, \citenamefont {Lai},\ and\ \citenamefont {Grebogi}}]{wang2014nonlinear}%
  \BibitemOpen
  \bibfield  {author} {\bibinfo {author} {\bibfnamefont {G.~L.}\ \bibnamefont {Wang}}, \bibinfo {author} {\bibfnamefont {L.}~\bibnamefont {Huang}}, \bibinfo {author} {\bibfnamefont {Y.~C.}\ \bibnamefont {Lai}},\ and\ \bibinfo {author} {\bibfnamefont {C.}~\bibnamefont {Grebogi}},\ }\href {https://doi.org/10.1103/PhysRevLett.112.110406} {\bibfield  {journal} {\bibinfo  {journal} {Phys. Rev. Lett.}\ }\textbf {\bibinfo {volume} {112}},\ \bibinfo {pages} {110406} (\bibinfo {year} {2014})}\BibitemShut {NoStop}%
\bibitem [{\citenamefont {Yang}\ \emph {et~al.}(2022)\citenamefont {Yang}, \citenamefont {Yin}, \citenamefont {Zhang},\ and\ \citenamefont {Nie}}]{yang2022macroscopic}%
  \BibitemOpen
  \bibfield  {author} {\bibinfo {author} {\bibfnamefont {X.}~\bibnamefont {Yang}}, \bibinfo {author} {\bibfnamefont {H.-J.}\ \bibnamefont {Yin}}, \bibinfo {author} {\bibfnamefont {F.}~\bibnamefont {Zhang}},\ and\ \bibinfo {author} {\bibfnamefont {J.}~\bibnamefont {Nie}},\ }\href@noop {} {\bibfield  {journal} {\bibinfo  {journal} {Laser Phys. Lett.}\ }\textbf {\bibinfo {volume} {20}},\ \bibinfo {pages} {015205} (\bibinfo {year} {2022})}\BibitemShut {NoStop}%
\bibitem [{\citenamefont {Walter}\ and\ \citenamefont {Marquardt}(2016)}]{walter2016classical}%
  \BibitemOpen
  \bibfield  {author} {\bibinfo {author} {\bibfnamefont {S.}~\bibnamefont {Walter}}\ and\ \bibinfo {author} {\bibfnamefont {F.}~\bibnamefont {Marquardt}},\ }\href {https://doi.org/10.1088/1367-2630/18/11/113029} {\bibfield  {journal} {\bibinfo  {journal} {New J. Phys.}\ }\textbf {\bibinfo {volume} {18}},\ \bibinfo {pages} {113029} (\bibinfo {year} {2016})}\BibitemShut {NoStop}%
\bibitem [{\citenamefont {L\"{u}}\ \emph {et~al.}(2015)\citenamefont {L\"{u}}, \citenamefont {Jing}, \citenamefont {Ma},\ and\ \citenamefont {Wu}}]{lu2015p}%
  \BibitemOpen
  \bibfield  {author} {\bibinfo {author} {\bibfnamefont {X.~Y.}\ \bibnamefont {L\"{u}}}, \bibinfo {author} {\bibfnamefont {H.}~\bibnamefont {Jing}}, \bibinfo {author} {\bibfnamefont {J.~Y.}\ \bibnamefont {Ma}},\ and\ \bibinfo {author} {\bibfnamefont {Y.}~\bibnamefont {Wu}},\ }\href {https://doi.org/10.1103/PhysRevLett.114.253601} {\bibfield  {journal} {\bibinfo  {journal} {Phys. Rev. Lett.}\ }\textbf {\bibinfo {volume} {114}},\ \bibinfo {pages} {253601} (\bibinfo {year} {2015})}\BibitemShut {NoStop}%
\bibitem [{\citenamefont {Sciamanna}(2016)}]{sciamanna2016vibrations}%
  \BibitemOpen
  \bibfield  {author} {\bibinfo {author} {\bibfnamefont {M.}~\bibnamefont {Sciamanna}},\ }\href {https://doi.org/10.1038/nphoton.2016.67} {\bibfield  {journal} {\bibinfo  {journal} {Nat. Photon.}\ }\textbf {\bibinfo {volume} {10}},\ \bibinfo {pages} {366} (\bibinfo {year} {2016})}\BibitemShut {NoStop}%
\bibitem [{\citenamefont {Navarro-Urrios}\ \emph {et~al.}(2017)\citenamefont {Navarro-Urrios}, \citenamefont {Capuj}, \citenamefont {Colombano}, \citenamefont {Garcia}, \citenamefont {Sledzinska}, \citenamefont {Alzina}, \citenamefont {Griol}, \citenamefont {Martinez},\ and\ \citenamefont {Sotomayor-Torres}}]{navarro2017nonlinear}%
  \BibitemOpen
  \bibfield  {author} {\bibinfo {author} {\bibfnamefont {D.}~\bibnamefont {Navarro-Urrios}}, \bibinfo {author} {\bibfnamefont {N.~E.}\ \bibnamefont {Capuj}}, \bibinfo {author} {\bibfnamefont {M.~F.}\ \bibnamefont {Colombano}}, \bibinfo {author} {\bibfnamefont {P.~D.}\ \bibnamefont {Garcia}}, \bibinfo {author} {\bibfnamefont {M.}~\bibnamefont {Sledzinska}}, \bibinfo {author} {\bibfnamefont {F.}~\bibnamefont {Alzina}}, \bibinfo {author} {\bibfnamefont {A.}~\bibnamefont {Griol}}, \bibinfo {author} {\bibfnamefont {A.}~\bibnamefont {Martinez}},\ and\ \bibinfo {author} {\bibfnamefont {C.~M.}\ \bibnamefont {Sotomayor-Torres}},\ }\href@noop {} {\bibfield  {journal} {\bibinfo  {journal} {Nat. Commun.}\ }\textbf {\bibinfo {volume} {8}},\ \bibinfo {pages} {14965} (\bibinfo {year} {2017})}\BibitemShut {NoStop}%
\bibitem [{\citenamefont {Lee}\ \emph {et~al.}(2009)\citenamefont {Lee}, \citenamefont {Yang}, \citenamefont {Moon}, \citenamefont {Lee}, \citenamefont {Shim}, \citenamefont {Kim}, \citenamefont {Lee},\ and\ \citenamefont {An}}]{lee2009observation}%
  \BibitemOpen
  \bibfield  {author} {\bibinfo {author} {\bibfnamefont {S.~B.}\ \bibnamefont {Lee}}, \bibinfo {author} {\bibfnamefont {J.}~\bibnamefont {Yang}}, \bibinfo {author} {\bibfnamefont {S.}~\bibnamefont {Moon}}, \bibinfo {author} {\bibfnamefont {S.~Y.}\ \bibnamefont {Lee}}, \bibinfo {author} {\bibfnamefont {J.~B.}\ \bibnamefont {Shim}}, \bibinfo {author} {\bibfnamefont {S.~W.}\ \bibnamefont {Kim}}, \bibinfo {author} {\bibfnamefont {J.~H.}\ \bibnamefont {Lee}},\ and\ \bibinfo {author} {\bibfnamefont {K.}~\bibnamefont {An}},\ }\href {https://doi.org/10.1103/PhysRevLett.103.134101} {\bibfield  {journal} {\bibinfo  {journal} {Phys. Rev. Lett.}\ }\textbf {\bibinfo {volume} {103}},\ \bibinfo {pages} {134101} (\bibinfo {year} {2009})}\BibitemShut {NoStop}%
\bibitem [{\citenamefont {Sun}\ and\ \citenamefont {Sukhorukov}(2014)}]{sun2014chaotic}%
  \BibitemOpen
  \bibfield  {author} {\bibinfo {author} {\bibfnamefont {Y.}~\bibnamefont {Sun}}\ and\ \bibinfo {author} {\bibfnamefont {A.~A.}\ \bibnamefont {Sukhorukov}},\ }\href {https://doi.org/10.1364/OL.39.003543} {\bibfield  {journal} {\bibinfo  {journal} {Opt. Lett.}\ }\textbf {\bibinfo {volume} {39}},\ \bibinfo {pages} {3543} (\bibinfo {year} {2014})}\BibitemShut {NoStop}%
\bibitem [{\citenamefont {Piazza}\ and\ \citenamefont {Ritsch}(2015)}]{piazza2015self}%
  \BibitemOpen
  \bibfield  {author} {\bibinfo {author} {\bibfnamefont {F.}~\bibnamefont {Piazza}}\ and\ \bibinfo {author} {\bibfnamefont {H.}~\bibnamefont {Ritsch}},\ }\href {https://doi.org/10.1103/PhysRevLett.115.163601} {\bibfield  {journal} {\bibinfo  {journal} {Phys. Rev. Lett.}\ }\textbf {\bibinfo {volume} {115}},\ \bibinfo {pages} {163601} (\bibinfo {year} {2015})}\BibitemShut {NoStop}%
\bibitem [{\citenamefont {Zhang}\ \emph {et~al.}(2020)\citenamefont {Zhang}, \citenamefont {You},\ and\ \citenamefont {L{\"u}}}]{zhang2020intermittent}%
  \BibitemOpen
  \bibfield  {author} {\bibinfo {author} {\bibfnamefont {D.}~\bibnamefont {Zhang}}, \bibinfo {author} {\bibfnamefont {C.}~\bibnamefont {You}},\ and\ \bibinfo {author} {\bibfnamefont {X.}~\bibnamefont {L{\"u}}},\ }\href@noop {} {\bibfield  {journal} {\bibinfo  {journal} {Phys. Rev. A}\ }\textbf {\bibinfo {volume} {101}},\ \bibinfo {pages} {053851} (\bibinfo {year} {2020})}\BibitemShut {NoStop}%
\bibitem [{\citenamefont {Wang}\ \emph {et~al.}(2016)\citenamefont {Wang}, \citenamefont {Lai},\ and\ \citenamefont {Grebogi}}]{wang2016transient}%
  \BibitemOpen
  \bibfield  {author} {\bibinfo {author} {\bibfnamefont {G.~L.}\ \bibnamefont {Wang}}, \bibinfo {author} {\bibfnamefont {Y.~C.}\ \bibnamefont {Lai}},\ and\ \bibinfo {author} {\bibfnamefont {C.}~\bibnamefont {Grebogi}},\ }\href {https://doi.org/10.1038/srep35381} {\bibfield  {journal} {\bibinfo  {journal} {Sci. Rep.}\ }\textbf {\bibinfo {volume} {6}},\ \bibinfo {pages} {35381} (\bibinfo {year} {2016})}\BibitemShut {NoStop}%
\bibitem [{\citenamefont {Bakemeier}\ \emph {et~al.}(2015)\citenamefont {Bakemeier}, \citenamefont {Alvermann},\ and\ \citenamefont {Fehske}}]{bakemeier2015route}%
  \BibitemOpen
  \bibfield  {author} {\bibinfo {author} {\bibfnamefont {L.}~\bibnamefont {Bakemeier}}, \bibinfo {author} {\bibfnamefont {A.}~\bibnamefont {Alvermann}},\ and\ \bibinfo {author} {\bibfnamefont {H.}~\bibnamefont {Fehske}},\ }\href@noop {} {\bibfield  {journal} {\bibinfo  {journal} {Phys. Rev. Lett.}\ }\textbf {\bibinfo {volume} {114}},\ \bibinfo {pages} {013601} (\bibinfo {year} {2015})}\BibitemShut {NoStop}%
\bibitem [{\citenamefont {Nakamura}(1993)}]{nakamura1993new}%
  \BibitemOpen
  \bibfield  {author} {\bibinfo {author} {\bibfnamefont {K.}~\bibnamefont {Nakamura}},\ }\href@noop {} {\emph {\bibinfo {title} {Quantum Chaos: A New Paradigm of Nonlinear Dynamics}}}\ (\bibinfo  {publisher} {Cambridge University Press, Cambridge},\ \bibinfo {year} {1993})\BibitemShut {NoStop}%
\bibitem [{\citenamefont {Heller}(1984)}]{heller1984bound}%
  \BibitemOpen
  \bibfield  {author} {\bibinfo {author} {\bibfnamefont {E.~J.}\ \bibnamefont {Heller}},\ }\href@noop {} {\bibfield  {journal} {\bibinfo  {journal} {Phys. Rev. Lett.}\ }\textbf {\bibinfo {volume} {53}},\ \bibinfo {pages} {1515} (\bibinfo {year} {1984})}\BibitemShut {NoStop}%
\bibitem [{\citenamefont {Heller}(2018)}]{heller2018semiclassical}%
  \BibitemOpen
  \bibfield  {author} {\bibinfo {author} {\bibfnamefont {E.~J.}\ \bibnamefont {Heller}},\ }\href@noop {} {\emph {\bibinfo {title} {The semiclassical way to dynamics and spectroscopy}}}\ (\bibinfo  {publisher} {Princeton University Press},\ \bibinfo {year} {2018})\BibitemShut {NoStop}%
\bibitem [{\citenamefont {Gutzwiller}(1990)}]{gutzwiller1990chaos}%
  \BibitemOpen
  \bibfield  {author} {\bibinfo {author} {\bibfnamefont {M.~C.}\ \bibnamefont {Gutzwiller}},\ }\href@noop {} {\emph {\bibinfo {title} {Chaos in classical and quantum mechanics}}}\ (\bibinfo  {publisher} {Springer Verlag},\ \bibinfo {address} {New York},\ \bibinfo {year} {1990})\BibitemShut {NoStop}%
\bibitem [{\citenamefont {Ullmo}(2008)}]{ullmo2008many}%
  \BibitemOpen
  \bibfield  {author} {\bibinfo {author} {\bibfnamefont {D.}~\bibnamefont {Ullmo}},\ }\href@noop {} {\bibfield  {journal} {\bibinfo  {journal} {Rep. Prog. Phys.}\ }\textbf {\bibinfo {volume} {71}},\ \bibinfo {pages} {026001} (\bibinfo {year} {2008})}\BibitemShut {NoStop}%
\bibitem [{\citenamefont {Wright}\ and\ \citenamefont {Weaver}(2010)}]{wright2010new}%
  \BibitemOpen
  \bibfield  {author} {\bibinfo {author} {\bibfnamefont {M.}~\bibnamefont {Wright}}\ and\ \bibinfo {author} {\bibfnamefont {R.}~\bibnamefont {Weaver}},\ }\href@noop {} {\emph {\bibinfo {title} {New directions in linear acoustics and vibration: quantum chaos, random matrix theory and complexity}}}\ (\bibinfo  {publisher} {Cambridge University Press},\ \bibinfo {year} {2010})\BibitemShut {NoStop}%
\bibitem [{\citenamefont {Garc{\'\i}a-Mata}\ \emph {et~al.}(2022)\citenamefont {Garc{\'\i}a-Mata}, \citenamefont {Jalabert},\ and\ \citenamefont {Wisniacki}}]{garcia2022out}%
  \BibitemOpen
  \bibfield  {author} {\bibinfo {author} {\bibfnamefont {I.}~\bibnamefont {Garc{\'\i}a-Mata}}, \bibinfo {author} {\bibfnamefont {R.~A.}\ \bibnamefont {Jalabert}},\ and\ \bibinfo {author} {\bibfnamefont {D.~A.}\ \bibnamefont {Wisniacki}},\ }\href@noop {} {\bibfield  {journal} {\bibinfo  {journal} {arXiv preprint arXiv:2209.07965}\ } (\bibinfo {year} {2022})}\BibitemShut {NoStop}%
\bibitem [{\citenamefont {Novotn{\`y}}\ and\ \citenamefont {Str{\'a}nsk{\`y}}(2023)}]{novotny2023relative}%
  \BibitemOpen
  \bibfield  {author} {\bibinfo {author} {\bibfnamefont {J.}~\bibnamefont {Novotn{\`y}}}\ and\ \bibinfo {author} {\bibfnamefont {P.}~\bibnamefont {Str{\'a}nsk{\`y}}},\ }\href@noop {} {\bibfield  {journal} {\bibinfo  {journal} {Phys. Rev. E}\ }\textbf {\bibinfo {volume} {107}},\ \bibinfo {pages} {054220} (\bibinfo {year} {2023})}\BibitemShut {NoStop}%
\bibitem [{\citenamefont {Roque}\ \emph {et~al.}(2020)\citenamefont {Roque}, \citenamefont {Marquardt},\ and\ \citenamefont {Yevtushenko}}]{roque2020nonlinear}%
  \BibitemOpen
  \bibfield  {author} {\bibinfo {author} {\bibfnamefont {T.~F.}\ \bibnamefont {Roque}}, \bibinfo {author} {\bibfnamefont {F.}~\bibnamefont {Marquardt}},\ and\ \bibinfo {author} {\bibfnamefont {O.~M.}\ \bibnamefont {Yevtushenko}},\ }\href@noop {} {\bibfield  {journal} {\bibinfo  {journal} {New J. Phys.}\ }\textbf {\bibinfo {volume} {22}},\ \bibinfo {pages} {013049} (\bibinfo {year} {2020})}\BibitemShut {NoStop}%
\bibitem [{\citenamefont {Breuer}\ and\ \citenamefont {Petruccione}(2002)}]{breuer2002theory}%
  \BibitemOpen
  \bibfield  {author} {\bibinfo {author} {\bibfnamefont {H.}~\bibnamefont {Breuer}}\ and\ \bibinfo {author} {\bibfnamefont {F.}~\bibnamefont {Petruccione}},\ }\href@noop {} {\emph {\bibinfo {title} {The theory of open quantum systems}}}\ (\bibinfo  {publisher} {Oxford University Press, USA},\ \bibinfo {year} {2002})\BibitemShut {NoStop}%
\bibitem [{\citenamefont {Weiss}(2012)}]{weiss2012quantum}%
  \BibitemOpen
  \bibfield  {author} {\bibinfo {author} {\bibfnamefont {U.}~\bibnamefont {Weiss}},\ }\href@noop {} {\emph {\bibinfo {title} {Quantum dissipative systems}}}\ (\bibinfo  {publisher} {World Scientific},\ \bibinfo {year} {2012})\BibitemShut {NoStop}%
\bibitem [{\citenamefont {De~Vega}\ and\ \citenamefont {Alonso}(2017)}]{de2017dynamics}%
  \BibitemOpen
  \bibfield  {author} {\bibinfo {author} {\bibfnamefont {I.}~\bibnamefont {De~Vega}}\ and\ \bibinfo {author} {\bibfnamefont {D.}~\bibnamefont {Alonso}},\ }\href@noop {} {\bibfield  {journal} {\bibinfo  {journal} {Reviews of Modern Physics}\ }\textbf {\bibinfo {volume} {89}},\ \bibinfo {pages} {015001} (\bibinfo {year} {2017})}\BibitemShut {NoStop}%
\bibitem [{\citenamefont {Miki}\ \emph {et~al.}(2023)\citenamefont {Miki}, \citenamefont {Matsumoto}, \citenamefont {Matsumura}, \citenamefont {Shichijo}, \citenamefont {Sugiyama}, \citenamefont {Yamamoto},\ and\ \citenamefont {Yamamoto}}]{miki2023generating}%
  \BibitemOpen
  \bibfield  {author} {\bibinfo {author} {\bibfnamefont {D.}~\bibnamefont {Miki}}, \bibinfo {author} {\bibfnamefont {N.}~\bibnamefont {Matsumoto}}, \bibinfo {author} {\bibfnamefont {A.}~\bibnamefont {Matsumura}}, \bibinfo {author} {\bibfnamefont {T.}~\bibnamefont {Shichijo}}, \bibinfo {author} {\bibfnamefont {Y.}~\bibnamefont {Sugiyama}}, \bibinfo {author} {\bibfnamefont {K.}~\bibnamefont {Yamamoto}},\ and\ \bibinfo {author} {\bibfnamefont {N.}~\bibnamefont {Yamamoto}},\ }\href@noop {} {\bibfield  {journal} {\bibinfo  {journal} {Phys. Rev. A}\ }\textbf {\bibinfo {volume} {107}},\ \bibinfo {pages} {032410} (\bibinfo {year} {2023})}\BibitemShut {NoStop}%
\bibitem [{\citenamefont {Liu}\ \emph {et~al.}(2023)\citenamefont {Liu}, \citenamefont {Jiao}, \citenamefont {Li}, \citenamefont {Xu}, \citenamefont {He},\ and\ \citenamefont {Jing}}]{liu2023phase}%
  \BibitemOpen
  \bibfield  {author} {\bibinfo {author} {\bibfnamefont {J.~X.}\ \bibnamefont {Liu}}, \bibinfo {author} {\bibfnamefont {Y.~F.}\ \bibnamefont {Jiao}}, \bibinfo {author} {\bibfnamefont {Y.}~\bibnamefont {Li}}, \bibinfo {author} {\bibfnamefont {X.~W.}\ \bibnamefont {Xu}}, \bibinfo {author} {\bibfnamefont {Q.~Y.}\ \bibnamefont {He}},\ and\ \bibinfo {author} {\bibfnamefont {H.}~\bibnamefont {Jing}},\ }\href@noop {} {\bibfield  {journal} {\bibinfo  {journal} {Sci. China: Phys. Mech. Astron.}\ }\textbf {\bibinfo {volume} {66}},\ \bibinfo {pages} {230312} (\bibinfo {year} {2023})}\BibitemShut {NoStop}%
\bibitem [{\citenamefont {Strunz}\ \emph {et~al.}(1999)\citenamefont {Strunz}, \citenamefont {Di{\'o}si},\ and\ \citenamefont {Gisin}}]{strunz1999open}%
  \BibitemOpen
  \bibfield  {author} {\bibinfo {author} {\bibfnamefont {W.~T.}\ \bibnamefont {Strunz}}, \bibinfo {author} {\bibfnamefont {L.}~\bibnamefont {Di{\'o}si}},\ and\ \bibinfo {author} {\bibfnamefont {N.}~\bibnamefont {Gisin}},\ }\href@noop {} {\bibfield  {journal} {\bibinfo  {journal} {Phys. Rev. Lett.}\ }\textbf {\bibinfo {volume} {82}},\ \bibinfo {pages} {1801} (\bibinfo {year} {1999})}\BibitemShut {NoStop}%
\bibitem [{\citenamefont {Yu}\ \emph {et~al.}(1999)\citenamefont {Yu}, \citenamefont {Di{\'o}si}, \citenamefont {Gisin},\ and\ \citenamefont {Strunz}}]{yu1999non}%
  \BibitemOpen
  \bibfield  {author} {\bibinfo {author} {\bibfnamefont {T.}~\bibnamefont {Yu}}, \bibinfo {author} {\bibfnamefont {L.}~\bibnamefont {Di{\'o}si}}, \bibinfo {author} {\bibfnamefont {N.}~\bibnamefont {Gisin}},\ and\ \bibinfo {author} {\bibfnamefont {W.~T.}\ \bibnamefont {Strunz}},\ }\href@noop {} {\bibfield  {journal} {\bibinfo  {journal} {Phys. Rev. A}\ }\textbf {\bibinfo {volume} {60}},\ \bibinfo {pages} {91} (\bibinfo {year} {1999})}\BibitemShut {NoStop}%
\bibitem [{\citenamefont {Di{\'o}si}\ \emph {et~al.}(1998)\citenamefont {Di{\'o}si}, \citenamefont {Gisin},\ and\ \citenamefont {Strunz}}]{diosi1998non}%
  \BibitemOpen
  \bibfield  {author} {\bibinfo {author} {\bibfnamefont {L.}~\bibnamefont {Di{\'o}si}}, \bibinfo {author} {\bibfnamefont {N.}~\bibnamefont {Gisin}},\ and\ \bibinfo {author} {\bibfnamefont {W.~T.}\ \bibnamefont {Strunz}},\ }\href@noop {} {\bibfield  {journal} {\bibinfo  {journal} {Phys. Rev. A}\ }\textbf {\bibinfo {volume} {58}},\ \bibinfo {pages} {1699} (\bibinfo {year} {1998})}\BibitemShut {NoStop}%
\bibitem [{\citenamefont {Strunz}\ and\ \citenamefont {Yu}(2004)}]{strunz2004convolutionless}%
  \BibitemOpen
  \bibfield  {author} {\bibinfo {author} {\bibfnamefont {W.~T.}\ \bibnamefont {Strunz}}\ and\ \bibinfo {author} {\bibfnamefont {T.}~\bibnamefont {Yu}},\ }\href@noop {} {\bibfield  {journal} {\bibinfo  {journal} {Phys. Rev. A}\ }\textbf {\bibinfo {volume} {69}},\ \bibinfo {pages} {052115} (\bibinfo {year} {2004})}\BibitemShut {NoStop}%
\bibitem [{\citenamefont {Yu}(2004)}]{yu2004non}%
  \BibitemOpen
  \bibfield  {author} {\bibinfo {author} {\bibfnamefont {T.}~\bibnamefont {Yu}},\ }\href@noop {} {\bibfield  {journal} {\bibinfo  {journal} {Phys. Rev. A}\ }\textbf {\bibinfo {volume} {69}},\ \bibinfo {pages} {062107} (\bibinfo {year} {2004})}\BibitemShut {NoStop}%
\bibitem [{\citenamefont {Jing}\ and\ \citenamefont {Yu}(2010)}]{jing2010non}%
  \BibitemOpen
  \bibfield  {author} {\bibinfo {author} {\bibfnamefont {J.}~\bibnamefont {Jing}}\ and\ \bibinfo {author} {\bibfnamefont {T.}~\bibnamefont {Yu}},\ }\href@noop {} {\bibfield  {journal} {\bibinfo  {journal} {Phys. Rev. Lett.}\ }\textbf {\bibinfo {volume} {105}},\ \bibinfo {pages} {240403} (\bibinfo {year} {2010})}\BibitemShut {NoStop}%
\bibitem [{\citenamefont {Yang}\ \emph {et~al.}(2012)\citenamefont {Yang}, \citenamefont {Miao},\ and\ \citenamefont {Chen}}]{yang2012nonadiabatic}%
  \BibitemOpen
  \bibfield  {author} {\bibinfo {author} {\bibfnamefont {H.}~\bibnamefont {Yang}}, \bibinfo {author} {\bibfnamefont {H.}~\bibnamefont {Miao}},\ and\ \bibinfo {author} {\bibfnamefont {Y.}~\bibnamefont {Chen}},\ }\href@noop {} {\bibfield  {journal} {\bibinfo  {journal} {Phys. Rev. A}\ }\textbf {\bibinfo {volume} {85}},\ \bibinfo {pages} {040101(R)} (\bibinfo {year} {2012})}\BibitemShut {NoStop}%
\bibitem [{\citenamefont {Chen}\ \emph {et~al.}(2014)\citenamefont {Chen}, \citenamefont {You},\ and\ \citenamefont {Yu}}]{chen2014exact}%
  \BibitemOpen
  \bibfield  {author} {\bibinfo {author} {\bibfnamefont {Y.}~\bibnamefont {Chen}}, \bibinfo {author} {\bibfnamefont {J.~Q.}\ \bibnamefont {You}},\ and\ \bibinfo {author} {\bibfnamefont {T.}~\bibnamefont {Yu}},\ }\href@noop {} {\bibfield  {journal} {\bibinfo  {journal} {Phys. Rev. A}\ }\textbf {\bibinfo {volume} {90}},\ \bibinfo {pages} {052104} (\bibinfo {year} {2014})}\BibitemShut {NoStop}%
\bibitem [{\citenamefont {Xu}\ \emph {et~al.}(2014)\citenamefont {Xu}, \citenamefont {Zhao}, \citenamefont {Jing}, \citenamefont {Wu},\ and\ \citenamefont {Yu}}]{xu2014perturbation}%
  \BibitemOpen
  \bibfield  {author} {\bibinfo {author} {\bibfnamefont {J.}~\bibnamefont {Xu}}, \bibinfo {author} {\bibfnamefont {X.}~\bibnamefont {Zhao}}, \bibinfo {author} {\bibfnamefont {J.}~\bibnamefont {Jing}}, \bibinfo {author} {\bibfnamefont {L.~A.}\ \bibnamefont {Wu}},\ and\ \bibinfo {author} {\bibfnamefont {T.}~\bibnamefont {Yu}},\ }\href@noop {} {\bibfield  {journal} {\bibinfo  {journal} {J. Phys. A: Math. Theor.}\ }\textbf {\bibinfo {volume} {47}},\ \bibinfo {pages} {435301} (\bibinfo {year} {2014})}\BibitemShut {NoStop}%
\bibitem [{\citenamefont {Dalibard}\ \emph {et~al.}(1992)\citenamefont {Dalibard}, \citenamefont {Castin},\ and\ \citenamefont {M{\o}lmer}}]{dalibard1992wave}%
  \BibitemOpen
  \bibfield  {author} {\bibinfo {author} {\bibfnamefont {J.}~\bibnamefont {Dalibard}}, \bibinfo {author} {\bibfnamefont {Y.}~\bibnamefont {Castin}},\ and\ \bibinfo {author} {\bibfnamefont {K.}~\bibnamefont {M{\o}lmer}},\ }\href@noop {} {\bibfield  {journal} {\bibinfo  {journal} {Phys. Rev. Lett.}\ }\textbf {\bibinfo {volume} {68}},\ \bibinfo {pages} {580} (\bibinfo {year} {1992})}\BibitemShut {NoStop}%
\bibitem [{\citenamefont {Gisin}\ and\ \citenamefont {Percival}(1992)}]{gisin1992quantum}%
  \BibitemOpen
  \bibfield  {author} {\bibinfo {author} {\bibfnamefont {N.}~\bibnamefont {Gisin}}\ and\ \bibinfo {author} {\bibfnamefont {I.~C.}\ \bibnamefont {Percival}},\ }\href@noop {} {\bibfield  {journal} {\bibinfo  {journal} {J. Phys. A: Math. Gen.}\ }\textbf {\bibinfo {volume} {25}},\ \bibinfo {pages} {5677} (\bibinfo {year} {1992})}\BibitemShut {NoStop}%
\bibitem [{\citenamefont {Carmichael}(1993)}]{carmichael1993open}%
  \BibitemOpen
  \bibfield  {author} {\bibinfo {author} {\bibfnamefont {H.}~\bibnamefont {Carmichael}},\ }\href@noop {} {\emph {\bibinfo {title} {An Open Systems Approach to Quantum Optics Springer-Verlag}}}\ (\bibinfo  {publisher} {Springer Berlin, Heidelberg},\ \bibinfo {address} {Berlin},\ \bibinfo {year} {1993})\BibitemShut {NoStop}%
\bibitem [{\citenamefont {Wiseman}\ and\ \citenamefont {Milburn}(1993)}]{wiseman1993interpretation}%
  \BibitemOpen
  \bibfield  {author} {\bibinfo {author} {\bibfnamefont {H.~M.}\ \bibnamefont {Wiseman}}\ and\ \bibinfo {author} {\bibfnamefont {G.~J.}\ \bibnamefont {Milburn}},\ }\href@noop {} {\bibfield  {journal} {\bibinfo  {journal} {Phys. Rev. A}\ }\textbf {\bibinfo {volume} {47}},\ \bibinfo {pages} {1652} (\bibinfo {year} {1993})}\BibitemShut {NoStop}%
\bibitem [{\citenamefont {Plenio}\ and\ \citenamefont {Knight}(1998)}]{plenio1998quantum}%
  \BibitemOpen
  \bibfield  {author} {\bibinfo {author} {\bibfnamefont {M.~B.}\ \bibnamefont {Plenio}}\ and\ \bibinfo {author} {\bibfnamefont {P.~L.}\ \bibnamefont {Knight}},\ }\href@noop {} {\bibfield  {journal} {\bibinfo  {journal} {Rev. Mod. Phys.}\ }\textbf {\bibinfo {volume} {70}},\ \bibinfo {pages} {101} (\bibinfo {year} {1998})}\BibitemShut {NoStop}%
\bibitem [{\citenamefont {Liu}\ \emph {et~al.}(2011)\citenamefont {Liu}, \citenamefont {Li}, \citenamefont {Huang}, \citenamefont {Li}, \citenamefont {Guo}, \citenamefont {Laine}, \citenamefont {Breuer},\ and\ \citenamefont {Piilo}}]{liu2011experimental}%
  \BibitemOpen
  \bibfield  {author} {\bibinfo {author} {\bibfnamefont {B.~H.}\ \bibnamefont {Liu}}, \bibinfo {author} {\bibfnamefont {L.}~\bibnamefont {Li}}, \bibinfo {author} {\bibfnamefont {Y.~F.}\ \bibnamefont {Huang}}, \bibinfo {author} {\bibfnamefont {C.~F.}\ \bibnamefont {Li}}, \bibinfo {author} {\bibfnamefont {G.~C.}\ \bibnamefont {Guo}}, \bibinfo {author} {\bibfnamefont {E.~M.}\ \bibnamefont {Laine}}, \bibinfo {author} {\bibfnamefont {H.~P.}\ \bibnamefont {Breuer}},\ and\ \bibinfo {author} {\bibfnamefont {J.}~\bibnamefont {Piilo}},\ }\href@noop {} {\bibfield  {journal} {\bibinfo  {journal} {Nat. Phys.}\ }\textbf {\bibinfo {volume} {7}},\ \bibinfo {pages} {931} (\bibinfo {year} {2011})}\BibitemShut {NoStop}%
\bibitem [{\citenamefont {Marquardt}\ \emph {et~al.}(2006)\citenamefont {Marquardt}, \citenamefont {Harris},\ and\ \citenamefont {Girvin}}]{marquardt2006dynamical}%
  \BibitemOpen
  \bibfield  {author} {\bibinfo {author} {\bibfnamefont {F.}~\bibnamefont {Marquardt}}, \bibinfo {author} {\bibfnamefont {J.~G.~E.}\ \bibnamefont {Harris}},\ and\ \bibinfo {author} {\bibfnamefont {S.~M.}\ \bibnamefont {Girvin}},\ }\href@noop {} {\bibfield  {journal} {\bibinfo  {journal} {Phys. Rev. Lett.}\ }\textbf {\bibinfo {volume} {96}},\ \bibinfo {pages} {103901} (\bibinfo {year} {2006})}\BibitemShut {NoStop}%
\bibitem [{\citenamefont {Ludwig}\ \emph {et~al.}(2008)\citenamefont {Ludwig}, \citenamefont {Kubala},\ and\ \citenamefont {Marquardt}}]{ludwig2008optomechanical}%
  \BibitemOpen
  \bibfield  {author} {\bibinfo {author} {\bibfnamefont {M.}~\bibnamefont {Ludwig}}, \bibinfo {author} {\bibfnamefont {B.}~\bibnamefont {Kubala}},\ and\ \bibinfo {author} {\bibfnamefont {F.}~\bibnamefont {Marquardt}},\ }\href@noop {} {\bibfield  {journal} {\bibinfo  {journal} {New J. Phys.}\ }\textbf {\bibinfo {volume} {10}},\ \bibinfo {pages} {095013} (\bibinfo {year} {2008})}\BibitemShut {NoStop}%
\bibitem [{\citenamefont {Wolf}\ \emph {et~al.}(1985)\citenamefont {Wolf}, \citenamefont {Swift}, \citenamefont {Swinney},\ and\ \citenamefont {Vastano}}]{wolf1985determining}%
  \BibitemOpen
  \bibfield  {author} {\bibinfo {author} {\bibfnamefont {A.}~\bibnamefont {Wolf}}, \bibinfo {author} {\bibfnamefont {J.~B.}\ \bibnamefont {Swift}}, \bibinfo {author} {\bibfnamefont {H.~L.}\ \bibnamefont {Swinney}},\ and\ \bibinfo {author} {\bibfnamefont {J.~A.}\ \bibnamefont {Vastano}},\ }\href@noop {} {\bibfield  {journal} {\bibinfo  {journal} {Physica D}\ }\textbf {\bibinfo {volume} {16}},\ \bibinfo {pages} {285} (\bibinfo {year} {1985})}\BibitemShut {NoStop}%
\bibitem [{\citenamefont {Mu}\ \emph {et~al.}(2016)\citenamefont {Mu}, \citenamefont {Zhao},\ and\ \citenamefont {Yu}}]{mu2016memory}%
  \BibitemOpen
  \bibfield  {author} {\bibinfo {author} {\bibfnamefont {Q.}~\bibnamefont {Mu}}, \bibinfo {author} {\bibfnamefont {X.}~\bibnamefont {Zhao}},\ and\ \bibinfo {author} {\bibfnamefont {T.}~\bibnamefont {Yu}},\ }\href@noop {} {\bibfield  {journal} {\bibinfo  {journal} {Phys. Rev. A}\ }\textbf {\bibinfo {volume} {94}},\ \bibinfo {pages} {012334} (\bibinfo {year} {2016})}\BibitemShut {NoStop}%
\bibitem [{\citenamefont {Harayama}\ \emph {et~al.}(2011)\citenamefont {Harayama}, \citenamefont {Sunada}, \citenamefont {Yoshimura}, \citenamefont {Davis}, \citenamefont {Tsuzuki},\ and\ \citenamefont {Uchida}}]{harayama2011fast}%
  \BibitemOpen
  \bibfield  {author} {\bibinfo {author} {\bibfnamefont {T.}~\bibnamefont {Harayama}}, \bibinfo {author} {\bibfnamefont {S.}~\bibnamefont {Sunada}}, \bibinfo {author} {\bibfnamefont {K.}~\bibnamefont {Yoshimura}}, \bibinfo {author} {\bibfnamefont {P.}~\bibnamefont {Davis}}, \bibinfo {author} {\bibfnamefont {K.}~\bibnamefont {Tsuzuki}},\ and\ \bibinfo {author} {\bibfnamefont {A.}~\bibnamefont {Uchida}},\ }\href@noop {} {\bibfield  {journal} {\bibinfo  {journal} {Phys. Rev. A}\ }\textbf {\bibinfo {volume} {83}},\ \bibinfo {pages} {031803} (\bibinfo {year} {2011})}\BibitemShut {NoStop}%
\bibitem [{\citenamefont {Shen}\ \emph {et~al.}(2023)\citenamefont {Shen}, \citenamefont {Shu}, \citenamefont {Xie}, \citenamefont {Chen}, \citenamefont {Liu}, \citenamefont {Ge}, \citenamefont {Zhang}, \citenamefont {Wang}, \citenamefont {Zhang}, \citenamefont {Cheng} \emph {et~al.}}]{shen2023harnessing}%
  \BibitemOpen
  \bibfield  {author} {\bibinfo {author} {\bibfnamefont {B.}~\bibnamefont {Shen}}, \bibinfo {author} {\bibfnamefont {H.}~\bibnamefont {Shu}}, \bibinfo {author} {\bibfnamefont {W.}~\bibnamefont {Xie}}, \bibinfo {author} {\bibfnamefont {R.}~\bibnamefont {Chen}}, \bibinfo {author} {\bibfnamefont {Z.}~\bibnamefont {Liu}}, \bibinfo {author} {\bibfnamefont {Z.}~\bibnamefont {Ge}}, \bibinfo {author} {\bibfnamefont {X.}~\bibnamefont {Zhang}}, \bibinfo {author} {\bibfnamefont {Y.}~\bibnamefont {Wang}}, \bibinfo {author} {\bibfnamefont {Y.}~\bibnamefont {Zhang}}, \bibinfo {author} {\bibfnamefont {B.}~\bibnamefont {Cheng}}, \emph {et~al.},\ }\href@noop {} {\bibfield  {journal} {\bibinfo  {journal} {Nat. Commun.}\ }\textbf {\bibinfo {volume} {14}},\ \bibinfo {pages} {4590} (\bibinfo {year} {2023})}\BibitemShut {NoStop}%
\bibitem [{\citenamefont {Sun}\ \emph {et~al.}(2020)\citenamefont {Sun}, \citenamefont {Cai}, \citenamefont {Tang}, \citenamefont {Hu},\ and\ \citenamefont {Fan}}]{sun2020out}%
  \BibitemOpen
  \bibfield  {author} {\bibinfo {author} {\bibfnamefont {Z.-H.}\ \bibnamefont {Sun}}, \bibinfo {author} {\bibfnamefont {J.-Q.}\ \bibnamefont {Cai}}, \bibinfo {author} {\bibfnamefont {Q.-C.}\ \bibnamefont {Tang}}, \bibinfo {author} {\bibfnamefont {Y.}~\bibnamefont {Hu}},\ and\ \bibinfo {author} {\bibfnamefont {H.}~\bibnamefont {Fan}},\ }\href@noop {} {\bibfield  {journal} {\bibinfo  {journal} {Annalen der Physik}\ }\textbf {\bibinfo {volume} {532}},\ \bibinfo {pages} {1900270} (\bibinfo {year} {2020})}\BibitemShut {NoStop}%
\bibitem [{\citenamefont {Ch{\'a}vez-Carlos}\ \emph {et~al.}(2019)\citenamefont {Ch{\'a}vez-Carlos}, \citenamefont {L{\'o}pez-del Carpio}, \citenamefont {Bastarrachea-Magnani}, \citenamefont {Str{\'a}nsk{\`y}}, \citenamefont {Lerma-Hern{\'a}ndez}, \citenamefont {Santos},\ and\ \citenamefont {Hirsch}}]{chavez2019quantum}%
  \BibitemOpen
  \bibfield  {author} {\bibinfo {author} {\bibfnamefont {J.}~\bibnamefont {Ch{\'a}vez-Carlos}}, \bibinfo {author} {\bibfnamefont {B.}~\bibnamefont {L{\'o}pez-del Carpio}}, \bibinfo {author} {\bibfnamefont {M.~A.}\ \bibnamefont {Bastarrachea-Magnani}}, \bibinfo {author} {\bibfnamefont {P.}~\bibnamefont {Str{\'a}nsk{\`y}}}, \bibinfo {author} {\bibfnamefont {S.}~\bibnamefont {Lerma-Hern{\'a}ndez}}, \bibinfo {author} {\bibfnamefont {L.~F.}\ \bibnamefont {Santos}},\ and\ \bibinfo {author} {\bibfnamefont {J.~G.}\ \bibnamefont {Hirsch}},\ }\href@noop {} {\bibfield  {journal} {\bibinfo  {journal} {Phys. Rev. Lett.}\ }\textbf {\bibinfo {volume} {122}},\ \bibinfo {pages} {024101} (\bibinfo {year} {2019})}\BibitemShut {NoStop}%
\bibitem [{\citenamefont {Buijsman}\ \emph {et~al.}(2017)\citenamefont {Buijsman}, \citenamefont {Gritsev},\ and\ \citenamefont {Sprik}}]{buijsman2017nonergodicity}%
  \BibitemOpen
  \bibfield  {author} {\bibinfo {author} {\bibfnamefont {W.}~\bibnamefont {Buijsman}}, \bibinfo {author} {\bibfnamefont {V.}~\bibnamefont {Gritsev}},\ and\ \bibinfo {author} {\bibfnamefont {R.}~\bibnamefont {Sprik}},\ }\href@noop {} {\bibfield  {journal} {\bibinfo  {journal} {Phys. Rev. Lett.}\ }\textbf {\bibinfo {volume} {118}},\ \bibinfo {pages} {080601} (\bibinfo {year} {2017})}\BibitemShut {NoStop}%
\bibitem [{\citenamefont {Scala}\ \emph {et~al.}(2021)\citenamefont {Scala}, \citenamefont {S{\l}owik}, \citenamefont {Facchi}, \citenamefont {Pascazio},\ and\ \citenamefont {Pepe}}]{scala2021beyond}%
  \BibitemOpen
  \bibfield  {author} {\bibinfo {author} {\bibfnamefont {G.}~\bibnamefont {Scala}}, \bibinfo {author} {\bibfnamefont {K.}~\bibnamefont {S{\l}owik}}, \bibinfo {author} {\bibfnamefont {P.}~\bibnamefont {Facchi}}, \bibinfo {author} {\bibfnamefont {S.}~\bibnamefont {Pascazio}},\ and\ \bibinfo {author} {\bibfnamefont {F.~V.}\ \bibnamefont {Pepe}},\ }\href@noop {} {\bibfield  {journal} {\bibinfo  {journal} {Phys. Rev. A}\ }\textbf {\bibinfo {volume} {104}},\ \bibinfo {pages} {013722} (\bibinfo {year} {2021})}\BibitemShut {NoStop}%
\bibitem [{\citenamefont {Burgarth}\ \emph {et~al.}(2023)\citenamefont {Burgarth}, \citenamefont {Facchi}, \citenamefont {Hillier},\ and\ \citenamefont {Ligab{\`o}}}]{burgarth2023taming}%
  \BibitemOpen
  \bibfield  {author} {\bibinfo {author} {\bibfnamefont {D.}~\bibnamefont {Burgarth}}, \bibinfo {author} {\bibfnamefont {P.}~\bibnamefont {Facchi}}, \bibinfo {author} {\bibfnamefont {R.}~\bibnamefont {Hillier}},\ and\ \bibinfo {author} {\bibfnamefont {M.}~\bibnamefont {Ligab{\`o}}},\ }\href@noop {} {\bibfield  {journal} {\bibinfo  {journal} {arXiv preprint arXiv:2301.02269}\ } (\bibinfo {year} {2023})}\BibitemShut {NoStop}%
\end{thebibliography}%

\end{document}